\documentclass[twocolumn,aps,preprintnumbers,amsfonts,amsmath,amssymb,superscriptaddress,floatfix,10pt]{revtex4-2}

\usepackage{my-macro-equation}

\usepackage{bm}
 \usepackage{graphicx} 

\graphicspath{{./fig/}}

\newcommand\figref[1]{Fig.~\ref{fig:#1}}
\newcommand\figureref[1]{Figure~\ref{fig:#1}}
\newcommand\figuresref[1]{Figures~\ref{fig:#1}}

\usepackage[hidelinks]{hyperref} 
\begin{document}
\title{Feasibility study of a coherent feedback squeezer}

\author{Shota Yokoyama}
\email{s.yokoyama@unsw.edu.au}
\affiliation{Centre for Quantum Computation and Communication Technology and School of Engineering and Information Technology, The University of New South Wales, Canberra,  ACT 2600, Australia}

\author{Daniel Peace}
\affiliation{Centre for Quantum Computation and Communication Technology and Centre for Quantum Dynamics, Griffith University, Brisbane, QLD 4111, Australia}

\author{Warit Asavanant}
\affiliation{Department of Applied Physics, School of Engineering, The University of Tokyo,\\ 7-3-1 Hongo, Bunkyo-ku, Tokyo 113-8656, Japan}

\author{Takeyoshi Tajiri}
\affiliation{School of Engineering and Information Technology, The University of New South Wales, \\ Canberra, ACT 2600, Australia}
\affiliation{Institute of Industrial Science, The University of Tokyo, 4-6-1 Komaba, Meguro-ku, Tokyo 153-8505, Japan}

\author{Ben Haylock}
\affiliation{Centre for Quantum Computation and Communication Technology and Centre for Quantum Dynamics, Griffith University, Brisbane, QLD 4111, Australia}

\author{Moji Ghadimi}
\affiliation{Centre for Quantum Dynamics, Griffith University, Brisbane, QLD 4111, Australia}

\author{Mirko Lobino}
\affiliation{Centre for Quantum Computation and Communication Technology and Centre for Quantum Dynamics, Griffith University, Brisbane, QLD 4111, Australia}
\affiliation{Queensland Micro and Nanotechnology Centre, Griffith University, Brisbane, QLD 4111, Australia}

\author{Elanor H.\ Huntington}
\affiliation{Centre for Quantum Computation and Communication Technology and Research School of Electrical, Energy and Materials Engineering, College of Engineering and Computer Science, Australian National University, Canberra, ACT 2600, Australia}

\author{Hidehiro Yonezawa}
\email{h.yonezawa@unsw.edu.au}
\affiliation{Centre for Quantum Computation and Communication Technology and School of Engineering and Information Technology, The University of New South Wales, Canberra, ACT 2600, Australia}

\begin{abstract}
We investigate a coherent feedback squeezer that uses quantum coherent feedback (measurement-free) control. Our squeezer is simple, easy to implement, robust to the gain fluctuation, and broadband compared to the existing squeezers because of the negative coherent feedback configuration. We conduct a feasibility study that looks at the stability conditions for a feedback system to optimize the designs of real optical devices. The feasibility study gives fabrication tolerance necessary for  designing and realizing the actual device. Our formalism for the stability analysis is not limited to optical systems but can be applied to the other bosonic systems.
\end{abstract}

\maketitle
\section{Introduction}
A squeezer is one of the most fundamental components in continuous-variable (CV) quantum information processing \cite{Furusawa11}. One reason is that a squeezer generates a squeezed vacuum state whose uncertainty in one of two quadratures is squeezed less than that of a vacuum state. A squeezed vacuum state can be used as an ancilla in a wide range of CV protocols \cite{Yonezawa12,Lenzini18} to improve the sensitivity of metrology or capacity of quantum communication, and to generate quantum entanglement. Another reason of importance is that a squeezer realizes a squeezing operation which is built into almost all CV quantum state manipulation protocols. By combining a squeezer with passive linear optics, it can implement arbitrary Gaussian multi-mode transformations \cite{Braunstein05} and a quantum cubic phase gate which is one of the non-Gaussian one-mode transformations \cite{Gottesman01,Marek11}, leading to universal quantum computing \cite{Lloyd99}.

The most common implementation of a squeezer in the continuous-wave optical regime is a degenerate optical parametric oscillator (DOPO) consisting of a nonlinear crystal with a pump inside an optical cavity. This configuration has been the \textit{de facto} standard for generating squeezed vacuum states over the past 30 years \cite{Andersen16}. Currently, the highest squeezing level reported is about $-15$~dB \cite{Vahlbruch16}, which almost reaches the well-known thresholds ($-15$ to $-17$~dB) to realize fault-tolerance in measurement-based CV quantum computation \cite{Asavanant19,Walshe19}. However, DOPO has two drawbacks when using it to perform squeezing operation on an arbitrary input state other than a vacuum state: an optical loss inside the cavity degrades the input, and the controllability of the strength of squeezing operation has low precision since it depends on the pump power. Therefore, DOPO is not suitable to realize a squeezer for an arbitrary input state.

One of the other implementations is a measurement-and-feedforward squeezer (MFS) utilizing a squeezed vacuum state as an ancilla generated by DOPO \cite{Filip05,Yoshikawa07,Miwa14,Miyata14}. In MFS, an input state is combined with a squeezed vacuum state by a beamsplitter, and one of the outputs of the beamsplitter is measured by homodyne detection. According to the outcome of the homodyne detection, a displacement operation is implemented on the remaining output, resulting in a squeezing operation for the input. An input state does not propagate through the DOPO, avoiding the drastic degradation. The squeezing strength of the operation is determined by the beamsplitter reflectivity, which is often configured by the combination of a half wave plate and a polarization beamsplitter \cite{Yoshikawa07}. By rotating the half wave plate, the effective reflectivity is easily changeable, which leads to the precise controllability of the squeezing strength. However, the bandwidth of the squeezing operation is limited by the bandwidth of the electrical circuits. In addition, the electric gain of the feedforward (the displacement operation) must be adjusted according to the beamsplitter reflectivity, which makes the fast control of the feedforward gain difficult. For example, the bandwidth of a dynamic squeezing operation demonstrated in Ref.~\cite{Miyata14} is about 1~MHz. Furthermore, in MFS, we have to synchronize electric and optical signals for feedforward, which demands a long optical delay line (e.g., 12~m of optical delay line was used in Ref.~\cite{Miwa14}). Thus, MFS can realize a squeezer for an arbitrary input state, but there are experimental challenges to be solved.

\begin{figure}[b]
	\centering
	\includegraphics[scale=1.0, clip]{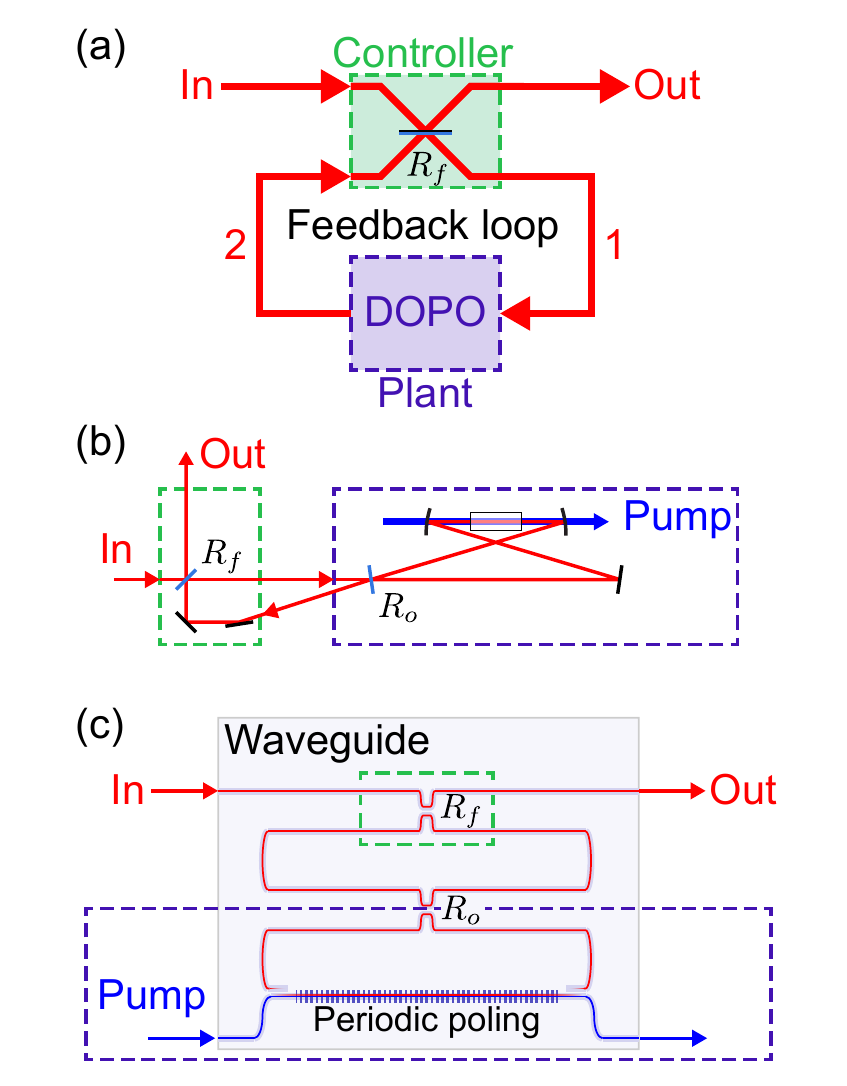}
	\caption{Coherent feedback squeezer. (a) The system consists of DOPO (plant) and a negative feedback loop by a beamsplitter (controller) with an energy reflectivity of $R_f$. Configurations in (b) free space and (c) optical waveguide. DOPO has a nonlinear crystal or periodic poling with a pump inside a cavity with an output coupler $R_o$.}
	\label{fig:QOPamp}
\end{figure}

In this paper, we investigate a coherent feedback squeezer (CFS) [\figref{QOPamp}(a)], which is an alternative implementation based on a quantum coherent feedback (measurement-free) control \cite{Iida12,Nurdin17,Dong10,Gough10,Gough09,Yamamoto14,Yokotera19,Yoshimura19,Shimazu19,Yamamoto16,Pan18,Yan11,Zhang17,Zhou15}. CFS consists of a high-gain DOPO and a negative coherent feedback loop with a beamsplitter whose reflectivity determines the strength of the squeezing operation as in MFS. In contrast to MFS, CFS does not have any quantum measurement and feedforward (i.e. electronics and optical delay line), which allows fast control of the strength of the squeezing operation only by changing the reflectivity of the beamsplitter. Furthermore, the negative feedback configuration improves the sensitivity to gain fluctuation of the DOPO, and makes the bandwidth broader compared to the incorporated DOPO. Although the feedback system has such advantages, it also incurs the possibility of making the system unstable. Thus, we need to carefully consider the feasibility of whether the feedback loop causes oscillation or not.

Quantum coherent feedback system have been eagerly investigated in terms of the application and the stability \cite{Gough10,Yamamoto16,Yokotera19,Yoshimura19,Shimazu19}. However, previous studies do not fully cover the stability analysis involving certain actual properties such as wavelength dispersion, phase-matching bandwidth, and periodic resonances of the feedback loop. For example, the linear quantum feedback networks in Ref.~\cite{Gough10} include our CFS, and those authors discussed the stability, but they did not consider such complicated experimental parameters. In order to design and implement an actual experimental setup, it is a significant step to investigate how these parameters affect the stability in a negative feedback system. We conduct feasibility studies for possible implementations of a quantum CFS in both free space [\figref{QOPamp}(b)] and optical waveguides [\figref{QOPamp}(c)], where we consider complex system parameters as we mentioned above. This analysis gives us new knowledge in the form of the tolerance to the cavity length mismatch.

We note that there are several investigations about similar coherent feedback system, which maybe confusing as to what differentiates ours. For example, the setup in Ref.~\cite{Iida12} is equivalent to ours while their feedback polarity is positive, which means the main property is to enhance the amplification of the small gain \cite{Gough10}. The drawback of the positive feedback is that it will be sensitive for gain fluctuations of the original amplifier. The fluctuation itself is also amplified and disturbs the output, which is opposite to one of the properties of a negative feedback system.

Recently, Shimazu and Yamamoto reported useful applications of negative coherent feedback systems including quantum versions of the differentiator, integrator, self-oscillator, and active filters \cite{Shimazu19}. Negative coherent feedback makes a system robust and broadband as an electrical circuit using well-known op-amps \cite{Yamamoto16}. Their system consists of a non-degenerate optical parametric oscillator and a beamsplitter with a negative feedback loop, which is also quite similar to our CFS but whose function is a phase-preserving quantum amplification. Our CFS would be a more basic component in terms that the phase-preserving quantum amplifier can be decomposed to two squeezers and two beamsplitters \cite{Braunstein05,Shiozawa18}.

We show the details of modeling our system and derivations in the appendixes. The representation does not use any assumptions except for the linear time-invariant (LTI) system with bosons. Therefore, it is a general approach that can be applied to other physical systems. Our results will contribute to the development of quantum control theory and the broad range of quantum technologies utilizing quantum coherent feedback control.

\section{Coherent Feedback Squeezer}
\label{sec:CFS}
In this section, we will explain the CFS as shown in \figref{QOPamp}(a). We first derive the input-output relation of the CFS in the ideal case without optical loss and discuss the sensitivity to the gain fluctuation of a DOPO, which is a standard figure of merit in the feedback system. Then, we calculate the frequency spectrum of CFS, which indicates that the negative feedback configuration makes a squeezer flat-gain and broadband compared to a DOPO. Finally, we consider the stability condition of our coherent feedback system and clarify the role of actual experimental parameters that are crucial when designing an CFS. Detailed derivations of equations can be found in the appendixes.

\subsection{Input-output Relation}
The CFS consists of a DOPO, a beamsplitter with an energy reflectivity $R_f$, and a negative feedback loop [\figref{QOPamp}(a)]. In the context of quantum feedback, the DOPO and the beamsplitter are called a plant (target system) and a controller, respectively. We assume the input-output relation of an ideal (lossless) DOPO as
\begin{align}
	\label{eq:DOPO_q}
	&\hat x_2 = G_x \hat x_1, \qquad
	\hat p_2 = G_p \hat p_1,\\
	&G_x = 1/G_p = G>0,
\end{align}
where $\hat x$ and $\hat p$ are canonical conjugate variables of quantized electromagnetic fields ($[\hat x, \hat p] = i/2$) which can be represented by annihilation and creation operators [$\hat x= (\hat a+\hat a^\dagger)/2,\ \hat p= (\hat a-\hat a^\dagger)/(2i), \ [\hat a,\hat a^\dagger ]=1$], the subscriptions denote input and output modes, and $G$ is a gain of DOPO.

The input-output relation of a beamsplitter is written as 
\begin{align}
	\begin{pmatrix}
		\hat x_{\mathrm{out}} \\
		\hat x_{1}
	\end{pmatrix}
	=
	\begin{pmatrix}
		\sqrt{R_f} & \sqrt{1-R_f} \\
		\sqrt{1-R_f} & -\sqrt{R_f}
	\end{pmatrix}
	\begin{pmatrix}
		\hat x_{\mathrm{in}} \\
		\hat x_{2}
	\end{pmatrix}, \\
	\begin{pmatrix}
		\hat p_{\mathrm{out}} \\
		\hat p_{1}
	\end{pmatrix}
	=
	\begin{pmatrix}
		\sqrt{R_f} & \sqrt{1-R_f} \\
		\sqrt{1-R_f} & -\sqrt{R_f}
	\end{pmatrix}
	\begin{pmatrix}
		\hat p_{\mathrm{in}} \\
		\hat p_{2}
	\end{pmatrix}.
\end{align}
Note that the minus sign of the matrix element of $-\sqrt{R_f}$ represents the negative feedback. The ``loop gain'', defined as the total gain around a feedback loop, is $-G_q\sqrt{R_f}$ ($q = x, p$). The feedback factor is also defined as $R_f$. From the above equations, we can derive the input-output relation as
\begin{align}
	&\hat x_{\mathrm{out}} = G_x^{(\mathrm{fb})} \hat x_{\mathrm{in}}, \qquad
	\hat p_{\mathrm{out}} = G_p^{(\mathrm{fb})} \hat p_{\mathrm{in}}, \\
\label{eq:GqFB_DC}
	&G_q^{(\mathrm{fb})} \equiv \frac{\sqrt{R_f}+G_q}{1+G_q\sqrt{R_f}}, \quad (q= x, p).
\end{align}
In the limit of $G_x\to \infty$ and  $G_p \to 0$  ($G\to\infty$), the gain of the system (i.e. ``transfer function'') can be approximated as
\begin{align}
	G_x^{(\mathrm{fb})}\approx 1/G_p^{(\mathrm{fb})}\approx 1/\sqrt{R_f}.
\end{align}
This shows that the squeezing operation is applied to the input. In addition, the squeezing level of the operation is determined only by the reflectivity of the beamsplitter $R_{f}$, which can be precisely controlled.

\subsection{Sensitivity to Gain Fluctuation of Plant}
One of the advantages of the CFS is its robustness against gain fluctuation in the plant (DOPO). The sensitivity of a quantum feedback system is defined by the normalized gain fluctuation at the output of the system due to the gain fluctuation of the plant, and can be expressed as \cite{Yamamoto16}
\begin{align}
	\mathcal{S}_{\mathrm{CFS}}&\equiv 
	\frac{\left||G_q^{(\mathrm{fb})}(G_q+\delta G_q)|- |G_q^{(\mathrm{fb})}(G_q)|\right|}{ |G_q^{(\mathrm{fb})}(G_q)|} \\
	&\le \frac{\left|G_q^{(\mathrm{fb})}(G_q+\delta G_q)- G_q^{(\mathrm{fb})}(G_q)\right|}{ |G_q^{(\mathrm{fb})}(G_q)|}.
\end{align}
In our CFS, the upper bound of the sensitivity is calculated as
\begin{align}
	\mathcal{S}_{\mathrm{CFS}} &\le \frac{1-R_f}{1+R_f + \sqrt{R_f}\left( G_{q}+1/G_{q} \right)}\frac{|\delta G_q|}{G_q} \\
	&\le \frac{1-R_f}{(1+\sqrt{R_f})^2}\frac{|\delta G_q|}{G_q} < \frac{|\delta G_q|}{G_q} = \mathcal{S}_{\mathrm{DOPO}}.
\end{align}
This upper bound is less than the sensitivity of the DOPO for any feedback factor $R_{f}$ and original amplifier gain $G_q$, which includes both squeezing $G_p<1$ and anti-squeezing $G_x>1$ cases.

\subsection{Frequency Response}
\begin{figure}[b]
	\centering
	\includegraphics[scale=1.0, clip]{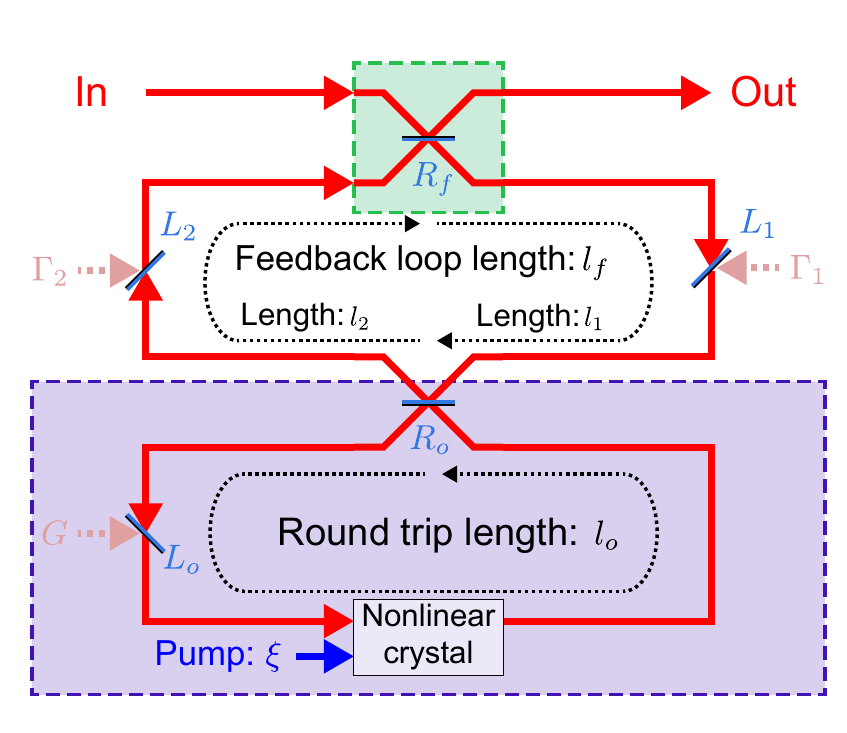}
	\caption{Schematics of CFS including realistic imperfections. $G$, $\Gamma_1$, and $\Gamma_2$ are unwanted noises at a DOPO, one before a DOPO, and one after a DOPO. Their optical losses and propagation length are $L_j$ and $l_j$ ($j=o,1,2$), respectively. $l_f = l_1+l_2$ is a total feedback loop length. $R_f$ and $R_o$ are the energy reflectivity of a beamsplitter at the feedback and DOPO output coupler. $\xi$ is a normalized pump amplitude.}
	\label{fig:CFS_parameters}
\end{figure}

The above discussion holds for all frequencies. However, to discuss bandwidth and stability, we must consider the frequency response of the CFS by using a realistic model with actual lengths,  propagation losses and DOPO  parameters [\figref{CFS_parameters}].

To do so, we first define the Fourier transform of an annihilation operator in the rotating frame of a carrier (angular) frequency $\omega_c$ as
	\begin{align}
	\label{eq:Fourier}
		\hat A (\omega ) \equiv \int_{-\infty}^{\infty} \left[\hat a(t)\ee^{i\omega_c t}\right]\ee ^{i \omega t}\dd t,
	\end{align}
which has canonical commutation relations as $[\hat A (\omega ), \hat A^\dagger (\omega')] = \delta (\omega - \omega')$ and $[\hat A (\omega ), \hat A (\omega')] = [\hat A^\dagger (\omega ), \hat A^\dagger (\omega')] = 0$, where $\delta (\cdot)$ is a Dirac delta function. We also define the Fourier transform of quadratures as 
	\begin{align}
	\label{eq:DefXomega}
		\hat X (\omega) & \equiv \frac{1}{2}\left[\hat A(\omega) + \hat A^\dagger(-\omega) \right], \\
	\label{eq:DefPomega}
		\hat P (\omega) & \equiv \frac{1}{2i}\left[\hat A(\omega) - \hat A^\dagger(-\omega) \right].
	\end{align}
In the form of the Fourier transform of quadratures, the input-output relation as a frequency function in CFS can be expressed as follows [see App.~\ref{sec:App_Frequency} for the derivation]:
	\begin{align}
	\notag
		&\hat X_{\mathrm{out}} (\omega ) = 
		G^{(\mathrm{CFS})}_{x}(\omega ) \hat X_{\mathrm{in}} (\omega ) 
		+ \sum _{j}G^{(\mathrm{CFS})}_{jx}(\omega ) \hat{X}_j (\omega), \\
	\notag
		&\hat P_{\mathrm{out}} (\omega ) = 
		G^{(\mathrm{CFS})}_{p}(\omega ) \hat P_{\mathrm{in}} (\omega ) 
		+ \sum _{j}G^{(\mathrm{CFS})}_{jp}(\omega ) \hat{P}_j (\omega),
	\end{align}
where $\hat{X}_j (\omega)$ and $\hat{P}_j (\omega)$ $(j= G, \Gamma_1, \Gamma_2)$ are the quadratures representing unwanted noises from the environment at the DOPO, one before the DOPO, and one after the DOPO, respectively [\figref{CFS_parameters}]. The transfer functions for an input signal and noises of the CFS are written as
\begin{align}
\notag
	G_q^{(\mathrm{CFS})}(\omega) & \equiv 
\frac{\sqrt{R_{f}}-\Lambda_q^{(\mathrm{CFS})}(\omega)/\sqrt{R_{f}}}{1-\Lambda_q^{(\mathrm{CFS})}(\omega)}, \\
\notag
	G_{G q}^{(\mathrm{CFS})}(\omega) & \equiv
\frac{\sqrt{(1-R_{f})(1-L_{2})}\ee^{i\omega\tau_2}}{1-\Lambda_q^{(\mathrm{CFS})}(\omega)}G_{\Delta q}^{(\mathrm{DOPO})}(\omega ),\\
\notag
	G_{\Gamma_1 q}^{(\mathrm{CFS})}(\omega) & \equiv 
\frac{\sqrt{(1-R_{f})L_{1}(1-L_2)}\ee^{i\omega\tau_2}}{1-\Lambda_q^{(\mathrm{CFS})}(\omega)}G_q^{(\mathrm{DOPO})}(\omega),\\
	G_{\Gamma_2 q}^{(\mathrm{CFS})}(\omega) & \equiv 
\frac{\sqrt{(1-R_{f})L_{2}}}{1-\Lambda_q^{(\mathrm{CFS})}(\omega)}, \quad (q=x,p),
\end{align}
which include the loop gain of CFS $\Lambda_q^{(\mathrm{CFS})}(\omega )$,
\begin{align}
	\Lambda_q^{(\mathrm{CFS})}(\omega ) &\equiv -\sqrt{R_{f}(1-L_{f})}\ee^{i\omega\tau_{f}} G^{(\mathrm{DOPO})}_q(\omega ),
\end{align}
and the transfer functions for an input signal and an external noise of the DOPO,
\begin{align}
\notag
	&G^{(\mathrm{DOPO})}_q(\omega ) \equiv \frac{(\gamma_{T_o} - \gamma_{L_o})/(2 \gamma ) + i\omega/\gamma +s_q\xi}{1- i\omega/\gamma -s_q\xi}, \\
\label{eq:TF_LangevinEq}
	&G_{\Delta q}^{(\mathrm{DOPO})} \equiv \frac{\sqrt{\gamma_{T_o}\gamma_{L_o}}/\gamma}{1-i\omega/\gamma  - s_q\xi}, \quad (s_x = -s_p = 1).
\end{align}
Here, we use many experimental parameters: $L_1\ (L_2)$, the optical loss at one after the DOPO with the length $l_1$ (one after the DOPO with the length $l_2$); $\tau_1 = l_1/c\  (\tau_2= l_2/c)$, the propagation time at $l_1\ (l_2)$; $c$, the speed of light in vacuum; $l_f = l_1+l_2$, the total feedback loop length; $L_f=1-(1-L_1)(1-L_2)$, the total propagation loss; $\tau _f = \tau_1 + \tau_2 $, the total propagation time; $T_o(=1-R_o)$, the energy transmittance of the DOPO output coupler; $L_o$ the intra-cavity loss of the DOPO; $l_o$, the round trip length of the DOPO; $\gamma= (\gamma_{T_o}+ \gamma_{L_o})/2$, decay rate at the DOPO where $\gamma_{T_o} = cT_o/l_o$ and $\gamma_{L_o} = cL_o/l_o$; and $\xi$, the pump amplitude normalized by the oscillation threshold amplitude ($0\le \xi <1$).

Note that the loop gain $\Lambda_q^{(\mathrm{CFS})}(\omega )$ for the quadrature $q$ is the total gain around a feedback loop, which consists of the effective feedback factor $R_f(1-L_f)$, the phase factor of the propagation $\ee^{i\omega \tau_f}$, and the DOPO gain $G_q^{(\mathrm{DOPO})} (\omega)$. In the case without propagation loss $L_f=0$ at the carrier frequency $\omega = 0$, the transfer function for an input signal $G_q^{(\mathrm{CFS})}(\omega = 0)$ reduces to Eq.~\eqref{eq:GqFB_DC}. In the ideal case without any losses, the output does not have any noise term, which means the operation is an unitary transformation on an input. On the other hand, MFS always has finite noise coming from an ancillary squeezed vacuum even in lossless case unless the ancillary state is infinitely squeezed. In that sense, CFS would be better than MFS in term of the final state purity.

\begin{figure}[t]
	\centering
	\includegraphics[scale=0.6, clip]{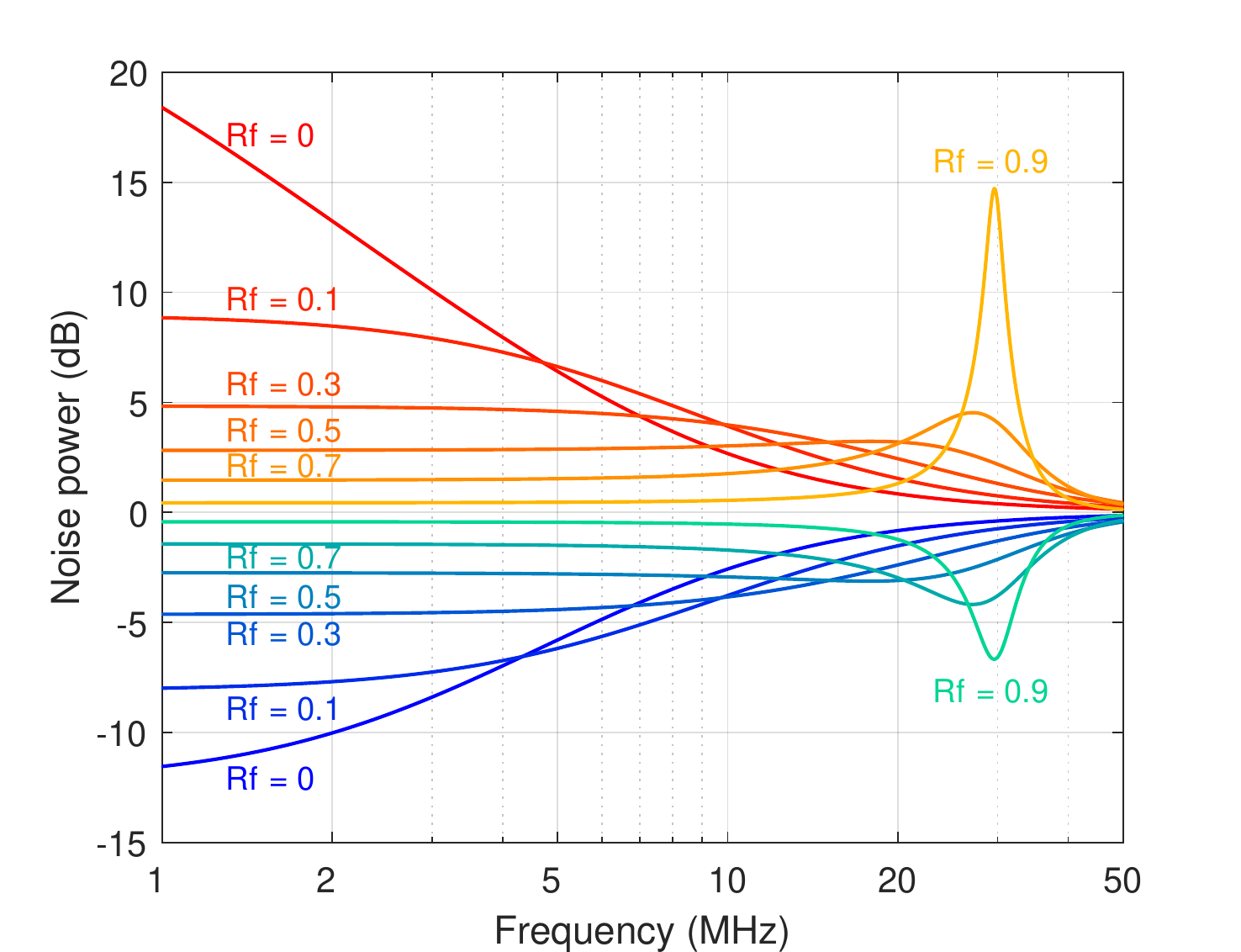}
	\caption{The output spectrums of CFS for vacuum input with several gains $R_{f}=0,0.1,0.3,0.5,0.7,0.9$. The power spectrums are normalized by vacuum fluctuation.}
	\label{fig:SpectrumCFS}
\end{figure}

\figureref{SpectrumCFS} shows simulations of the anti-squeezing and squeezing spectrums at the output of CFS for the vacuum input. Parameters used in the simulations are listed in Table~\ref{tb:SpectrumParam}, and are based on actual experimental parameters \cite{Takeno07}. The output power spectrums are normalized by the vacuum fluctuation. The traces with $R_{f}=0$ show the results without feedback, that is, the spectrums of the DOPO. As the feedback factor $R_{f}$ increases, the squeezing level decreases around the carrier frequency. As with a classical electrical amplifier consisting of an op-amp with negative feedback, the lower the squeezing level becomes (the stronger the feedback factor becomes), the wider the bandwidth becomes. However, if the feedback factor is too high, the gain peaking occurs due to the phase delay, which is observed at around 30~MHz with $R_{f}=0.9$. Note that the phase delay by the feedback loop may invert the feedback polarity, that is, in CFS, the squeezing gain is suppressed by the negative feedback around the carrier frequency while the gain may be increased at a higher frequency due to the phase delay. Therefore the bandwidth of the negative feedback system can be wider than that of the DOPO. We also note that, in a positive feedback system \cite{Iida12}, the gain is suppressed at a higher frequency due to the phase delay. Thus, the bandwidth of a positive feedback system is narrower than that of the DOPO.

In the classical op-amp design, when realizing such a low gain amplifier by a high feedback factor, we can avoid the gain peaking of the output by, for example, adding a capacitor in the feedback loop for phase compensation, or replacing the amplifier itself into unity-gain stable one. If we want to design CFS with the low squeezing, we can think of several possible methods such as lowering the gain of DOPO, reducing the phase delay of the feedback loop (designing the shorter loop), or adding a frequency filter in the loop. However, the problem is more complicated than the classical op-amp case. We will discuss the details of the stability condition in the next section.

\begin{table}
\centering
\caption{Parameters for CFS spectrum calculation in \figref{SpectrumCFS}.}
\label{tb:SpectrumParam}
\begin{tabular}{lcc}
\hline\hline
System parameter & Symbol & Value \\ \hline
Normalized pump amplitude & $\xi$ & 0.9 \\
DOPO output coupler & $T_o(=1-R_o)$ & 0.1 \\
DOPO intra-cavity loss & $L_o$ & 0.5\% \\
DOPO round trip length & $l_o$ & 500~mm \\
Controller BS & $R_{f}$ & 0, 0.1, 0.3 \\
			&   &  0.5, 0.7, 0.9 \\
Feedback loop loss & $L_{f} $ & 2\% \\
 & $L_1=L_2 $ & 1.01\% \\
Feedback loop length & $l_{f}$ & 500~mm \\ 
 & $l_1=l_2$ & 250~mm \\ \hline \hline
\end{tabular}
\end{table}

\subsection{Stability}
\label{sec:CFS_Stability}
The stability of the linear quantum system is defined as with classical linear systems \cite{Nurdin17}: \textit{We consider the quantum expectation value of a quadrature in a system as a function of time $\langle x(t) \rangle$. ``The system is said to be:
\begin{itemize}
\item Asymptotically stable or simply stable if $|\langle x(t) \rangle|\to 0$ as $t\to \infty$ for any initial state of the system.
\item Marginally stable if $|\langle x(t) \rangle|$ does not go to 0 as $t\to \infty$, but remains bounded at all times $t\ge 0$ for any initial state of the system.
\item Unstable if there exists some initial state of the system such that $|\langle x(t) \rangle|\to \infty$ as $t\to \infty$.''
\end{itemize}}
The stability analysis in quantum LTI systems can be conducted in the Laplace domain \cite{Yamamoto16,Yokotera19,Yoshimura19,Shimazu19}, that is, substituting $is$ into $\omega$ in a transfer function.
In order for a quantum LTI system to be stable, all of the roots of the characteristic equation,
	\begin{align}
		\Lambda_D(\omega = is) = 0,
	\end{align}
must have negative real parts. In our case, the characteristic function $\Lambda_D$ is written as [Eq.~\eqref{eq:KDquadratures}],
\begin{align}
	\Lambda_D(\omega) = [1-\Lambda_x(\omega )][1-\Lambda_p(\omega )]  
	 - \Lambda_{x\to p}(\omega )\Lambda_{p\to x}(\omega ),
\end{align}
where $\Lambda_x$, $\Lambda_p$, $\Lambda_{x\to p}$, and $\Lambda_{p\to x}$ are the elements of the loop gain matrix for the quadratures [Eq.~\eqref{eq:Gmatrix}]. The significant fact is that the characteristic equation consists of only the loop gains. Thus, it is enough to investigate the loop gains for the stability analysis. Note that, if we assume that the system has no wavelength dispersion, its symmetry around the carrier frequency [Eq.~\eqref{eq:DefSymmetry}] leads to $\Lambda_{x\to p}(\omega)\Lambda_{p\to x}(\omega)$ becoming zero [Eq.~\eqref{eq:Dsymmetry}].
 
Now, let us consider the stability of our CFS. In the case of an optical cavity, since there is a resonance every free spectral range (FSR, $c/l_o=0.6$~GHz from the parameters in Table~\ref{tb:SpectrumParam}), we need to consider not only the gains around the squeezing bandwidth (cavity line width, $\gamma/(2\pi)=5$~MHz from the parameters in Table~\ref{tb:SpectrumParam}), but also whole gains within the phase-matching bandwidth in the nonlinear material, which is typically of the order of THz and determined by the wavelength dispersion of the refractive index \cite{Gerrits11}. The DOPO transfer function shown in Eqs.~\eqref{eq:TF_LangevinEq} is derived from the quantum Langevin equation \cite{Collett84} considering only frequencies around the first resonance. This approximation does not account for the effect of length mismatch. In order to conduct a proper stability analysis, we need to replace Eqs~\eqref{eq:TF_LangevinEq} with other equations that incorporate the phase-matching bandwidth of the squeezing. We derived such a transfer function of the DOPO as $G_q^{(o,f)}(\omega)$ [Eq.~\eqref{eq:Gxof}] in free space, and the loop gain $\Lambda_q^{(\mathrm{CFS},f)} (\omega )$ [Eq.~\eqref{eq:KxCFSf}].

\begin{figure}
	\centering
	\includegraphics[scale=0.6, clip]{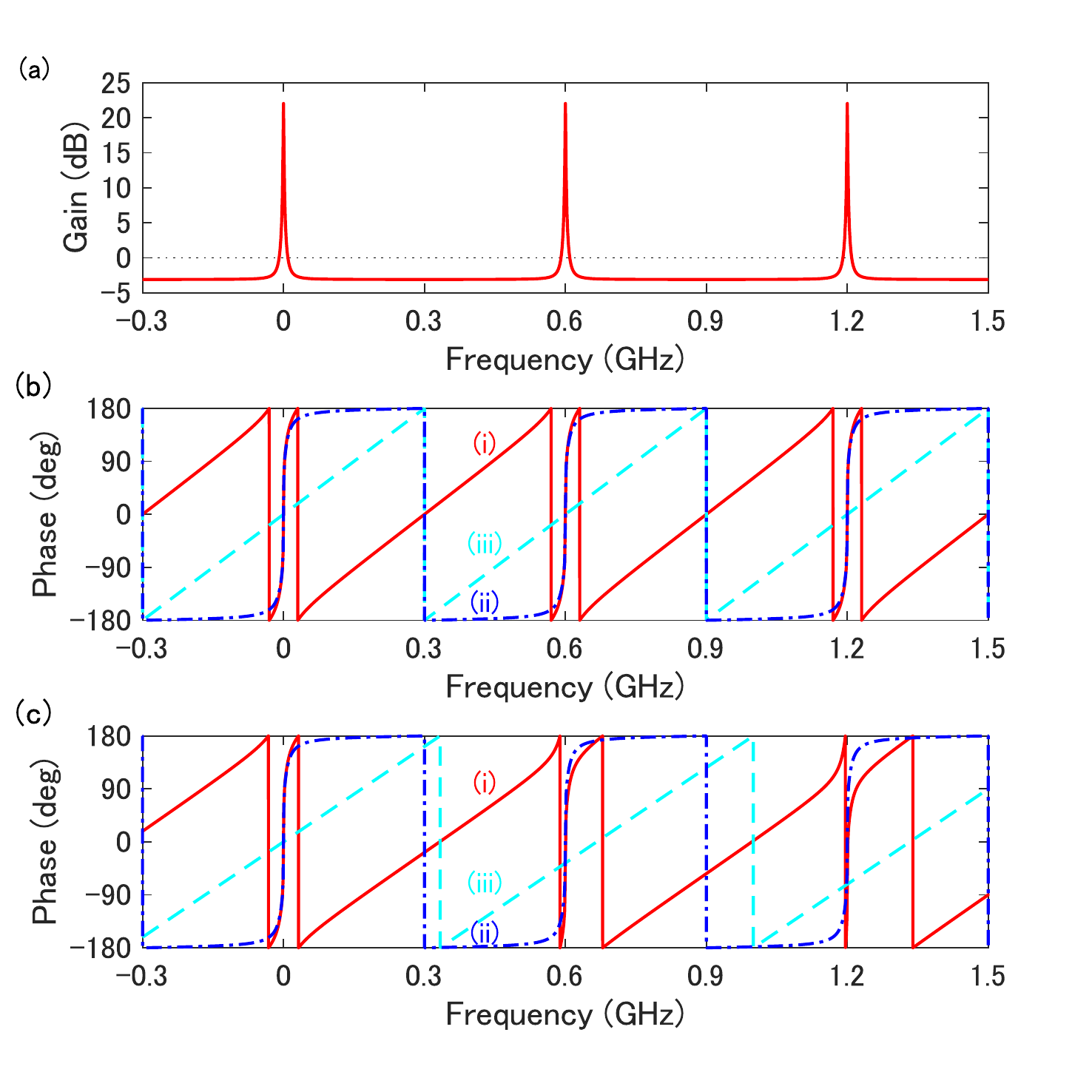}
	\includegraphics[scale=0.6, clip]{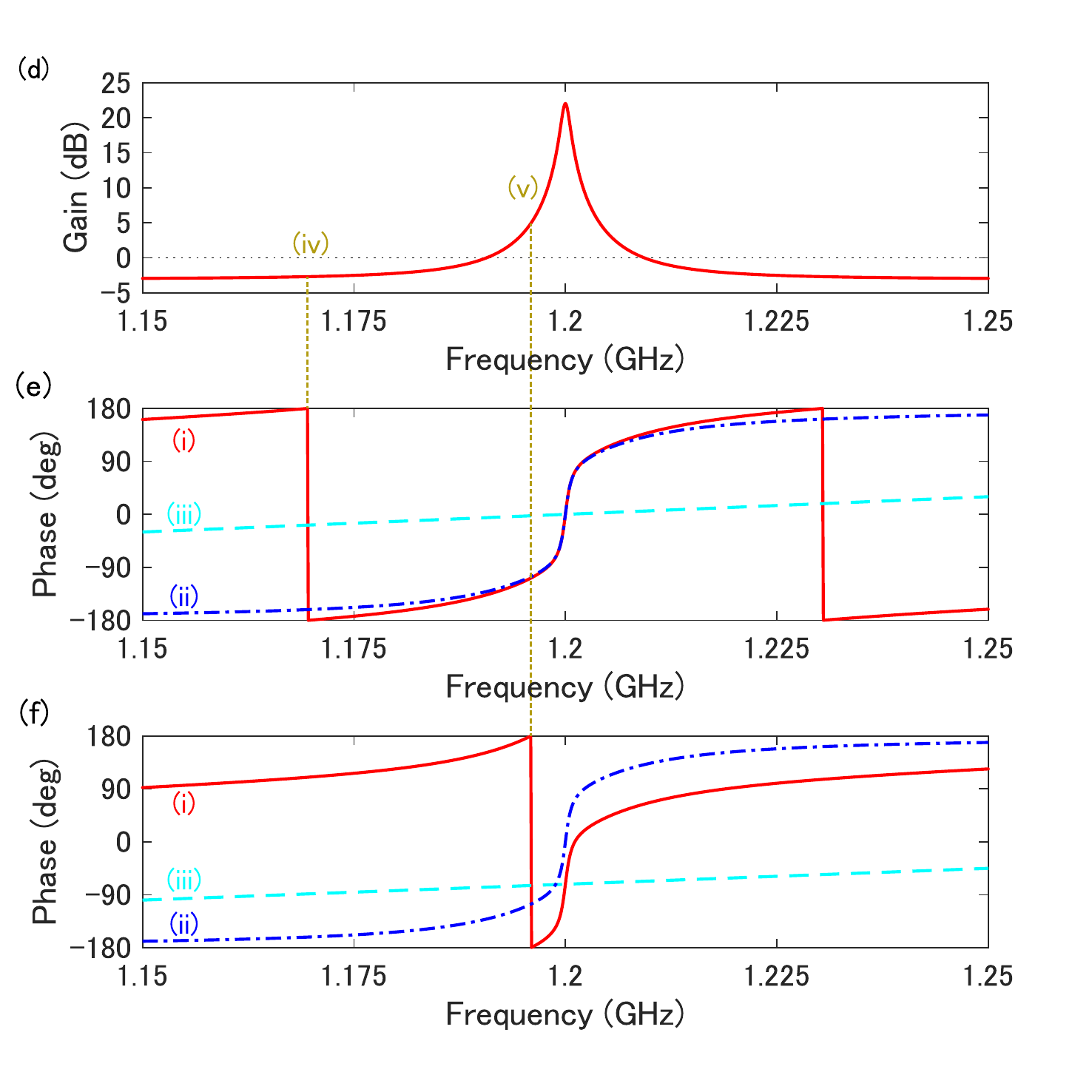}
	\caption{Bode plots of the loop gains of the feedback of CFS with free space optics. The loop gain $\Lambda_x^{(\mathrm{CFS},f)}(\omega)$ is given by Eq.~\eqref{eq:KxCFSf}. (a)(d) Gain $20\log_{10}|\Lambda_x^{(\mathrm{CFS},f)}(\omega)|$ with $l_{o}=500$~mm. ($R_{f}=0.5$, other parameters listed in Table~\ref{tb:SpectrumParam}). Phases in the case of (b)(e) $l_{f}=500$~mm and (c)(f) $l_{f}=450$~mm. (i) The phase of loop gain $\mathrm{arg} [-\Lambda_x^{(\mathrm{CFS},f)}(\omega)]$ is the sum of  (ii) the phase of feedback loop $\mathrm{arg} [\ee^{i \omega l_{f}/c}]$ and (iii) the phase of plant $\mathrm{arg} [G_x^{(o,f)}(\omega)]$ [Eq.~\eqref{eq:Gxof}]. The zero phase means that the feedback polarity is negative, while $\pm 180$~degree corresponds to the positive feedback. The loop gain $<$ 0~dB at the phase of $\pm 180$~degree is necessary for the system to be stable (iv), while the system oscillates if the loop gain $>$ 0~dB at the phase of $\pm 180$ degree (v). }
	\label{fig:LoopGain}
\end{figure}

\figuresref{LoopGain}(a)(b)(d)(e) show the Bode plot (the absolute value and the phase) of the loop gain of the feedback $-\Lambda_x^{(\mathrm{CFS},f)} (\omega )$ for the parameters given in Table~\ref{tb:SpectrumParam} and $R_{f}=0.5$. Here, we plot them up to the third resonance. The phase of the loop gain $\mathrm{arg} [-\Lambda_x^{(\mathrm{CFS},f)}(\omega)]$ [red trace (i) in (b) and (e)] is the sum of the phase of the DOPO $\mathrm{arg}[G_x^{(o,f)}(\omega)]$ [blue trace (ii) in (b) and (e)] and the phase of the feedback loop $\mathrm{arg}[\ee^{i\omega l_{f}/c}]$ [cyan trace (iii) in (b) and (e)]. Note that the zero phase means that the feedback polarity is negative, while $\pm 180$~degree corresponds to the positive feedback. The stability condition by a Bode plot is interpreted that the loop gain must be less than 1 ($=0$~dB) if the feedback polarity is positive (iv). The absolute value and the phase of loop gain have a periodic structure determined by FSR, while the period of the feedback loop phase is determined by the feedback loop length $l_{f}$. If the feedback loop length is the same as the DOPO round trip length ($l_{f}=l_{o}$), the periods are identical as \figref{LoopGain}(b)(e). \figuresref{LoopGain}(c)(f) show the phases in the case of $l_{f} =450$~mm ($\neq l_{o}$). Although the stability condition is satisfied within the first resonance, the third resonance has completely positive polarity due to the periods mismatching, which means the system is unstable (v). Thus, in addition to the feedback factor $R_{f}$ and the pump power, the length mismatch contributes to the stability condition.

\section{Feasibility study using stability condition}
In this section, we will conduct the stability analysis of CFS for both free space [\figref{QOPamp}(b)] and waveguide optics [\figref{QOPamp}(c)]. For the stability analysis, we consider the loop gains of feedback [App.~\ref{sec:LoopGains}] derived from the more general framework of a coherent feedback system [App.~\ref{sec:App_Theory}].

\subsection{CFS in Free Space}
\label{sec:CFSfmain}
Let us consider the design of a CFS with a DOPO built of free space optics as shown in \figref{QOPamp}(b). As we explained in Sec.~\ref{sec:CFS_Stability}, the stability condition is given in terms of  functions of several parameters: a pump amplitude normalized by the oscillation threshold amplitude $\xi$ ($0\le \xi <1$), a feedback factor $R_{f}$, and a length mismatch between the DOPO round trip length and the feedback loop length. Notably, because of the broad phase-matching bandwidth, the effect of the length mismatch is critical at high frequency. To reduce the sensitivity to the length mismatch, the narrower phase-matching bandwidth is better. Fortunately, in the case of free space optics, an optical bandpass filter (BPF) can be used to eliminate the higher frequency signals. In this analysis we assume the presence of a BPF with the cutoff frequency of 100~GHz in the feedback loop. The loop gain of CFS in free space is given as
\begin{align}
\notag
	\Lambda_q^{(\mathrm{CFS},f)}(\omega )& = -\sqrt{R_{f}(1-L_{f})} \ee^{i \omega l_{f}/c} \\
\label{eq:KxCFSf_main}
	&  \qquad \times H(\omega ) G_q^{(o,f)}(\omega ), \quad (q=x,p), 
\end{align}
where $H(\omega )$ is the transfer function of the BPF, $G_q^{(o,f)}(\omega )$ is the approximated transfer function at the DOPO around the carrier frequency [see App.~\ref{sec:CFSf} for the details]. 

To analyze the stability condition, we use the Nyquist plot which is the trajectory of the loop gain $-\Lambda_x^{{(\mathrm{CFS},f)}} (\omega )$ for all $\omega \in \mathbb{R}$ \cite{Yokotera19}. If and only if there is no encirclement of around $-1+0i$ in the complex plane, the system is said to be stable. We investigate the range of $\delta l_{f}$ ($l_{f}=l_{o} + \delta l_{f}$) that satisfies the stability condition for the parameters given in Table~\ref{tb:SpectrumParam} and $R_{f}\in [0,1)$, $\xi \in [0,1)$. We define the range as ``allowable length mismatch'' $= [\max (\delta l_{f})-\min (\delta l_{f})]/2$.

\begin{figure}[b]
	\centering
	\includegraphics[scale=0.6, clip]{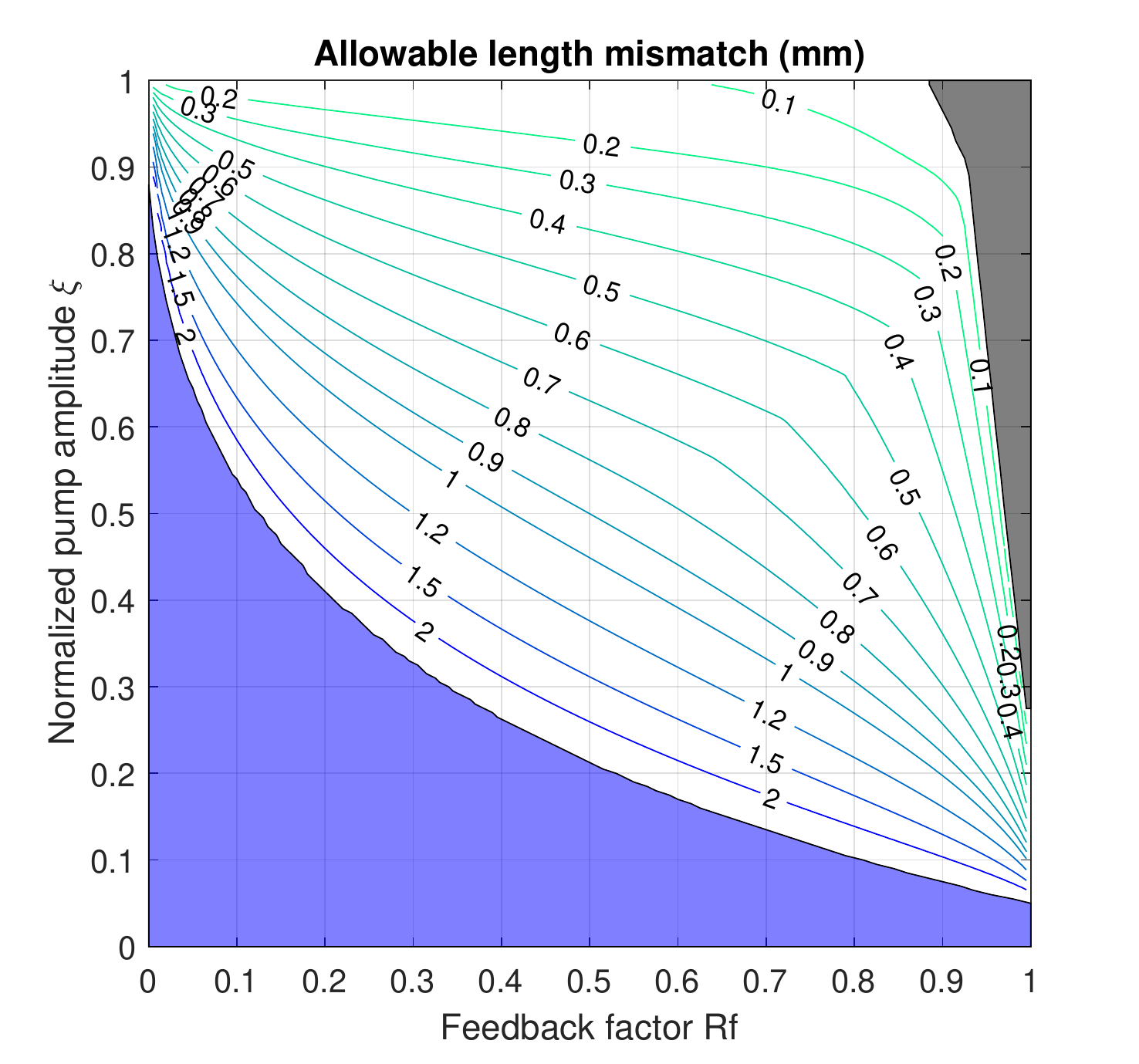}
	\caption{Allowable length mismatch to satisfy the stability condition about CFS with DOPO in free space optics. The blue-filled (gray-filled) area is the unconditionally stable (unstable) region, respectively.}
	\label{fig:Stability_DOPO}
\end{figure}

\figureref{Stability_DOPO} shows the calculated results of allowable length mismatch to satisfy the stability condition for a CFS made of free space optics. The horizontal and vertical axes correspond to the feedback factor $R_{f}$ and the normalized pump amplitude of the DOPO $\xi$, respectively. For parameters in the upper right, larger gains lead to instability in the system. The blue-filled area at the lower left is the unconditionally stable region, where there are no restrictions about $\delta l_{f}$ to satisfy stability condition because of the small gains. On the other hand, the grey-filled area at the top right is the unconditionally unstable region, which means the stability condition is never satisfied even if $\delta l_{f}=0$ because of the large feedback factor. In the middle area, the condition on the allowable length mismatch is obtained by numerical calculation. The line written as 0.1 means that the system works stably in the area of the lower left part of the line with 0.1~mm tolerance. The required precision to fabricate the device can be read from this figure.

For the full feasibility study, we may need to consider the other effects for the stability analysis based on various physics knowledge, which are not included in the above calculation. In the free space optics, the beam typically propagates as a Gaussian beam, which has a Gouy phase depending on the beam size and the propagation distance \cite{Yariv07}. Even if the two loop lengths are identical, the difference in beam size caused by the different cavity structures generates the phase difference. However, the difference can be compensated by adding an offset to the length, which can be negligible when calculated for a specific cavity design at a wavelength of 1550nm. In this particular design, the wavelength dispersion at the transmitting elements such as lens and mirrors changes the effective path length as long as the dispersion can be approximated as linear. At the stage of designing an experimental setup, we can adjust the length to compensate for the effect of those transmitting elements.

The required fabrication precision is a good indicator of what to ignore and what to consider. In this case, a precision of  0.1~mm is necessary in order to satisfy the stability condition in almost the entire region [\figref{Stability_DOPO}]. This value is crucial for the design of more reliable experimental setup.

\subsection{CFS in a Waveguide}
\label{sec:CFSwmain}
We now explore the design of CFS using waveguide integrated optics instead of free space optics. When compared with the free space implementation, with integrated optics each component can be arranged compactly on a single substrate, enabling large-scale integration and very-fast (broad bandwidth) operation. Even for this configuration, a stability analysis must be conducted on the effects of the wavelength dispersion in the waveguide medium and the broadband amplifier-gain bandwidth.

\figureref{QOPamp}(c) shows the schematics of CFS in a waveguide. We assume a ridge waveguide device using an $x$-cut 5~mol.~\% Magnesium-doped Lithium Niobate (MgO:LN) thin-film on an insulator substrate of SiO$_2$ \cite{Wang18}. The Lithium Niobate on an insulator (LNOI) platform, allows us to take advantage of the low propagation losses and high refractive index contrast enabling compact devices with small bending radii ($>100\mu$m) \cite{Zhang17-2,Krasnokutska18}. We use the effective refractive index of the waveguide $n_{\omega_c + \omega}$, which is calculated by using the Sellmeier equations for 5~mol.~\% MgO:LN \cite{Zelmon97} and for SiO$_2$ \cite{Malitson65}. High mode confinement of the LNOI platform results in geometry specific dispersion. Here we assume waveguide dimensions similar to that of Ref.~\cite{Zhang17-2}, which has demonstrated low propagation losses of 0.03~dB/cm. Short devices are favourable because they have less propagation losses. Nevertheless reducing the length broadens the phase-matching bandwidth and makes the system prones to oscillations. We suppose a periodically poled region of 5~mm ($= l_c$) and the round trip length as 11~mm ($=l_{o}$). The corresponding propagation losses are $L_{o}=L_{f}= 1-10^{-0.03\times1.1/10}=0.76$\%. To reduce the effect of intra-cavity loss, we suppose to use the higher reflectivity beamsplitter $R_{o} = 0.7$.

The loop gains of CFS in waveguide optics are derived with wavelength dispersion [see App.~~\ref{sec:CFSw} for the details]. The characteristic function $\Lambda_D^{(\mathrm{CFS},w)} (\omega )$ for the stability analysis consists of those loop gains. By substituting the above experimental parameters and given $R_{f}$ and $\xi$ into the characteristic equation $\Lambda_D^{(\mathrm{CFS},w)} (\omega ) =0$, we calculate the range of $\delta _{f}$ ($l_{f} = l_o +\delta _{f}$) to satisfy the stability condition as well as the free space optics case.

\begin{figure}[b]
	\centering
	\includegraphics[scale=0.6, clip]{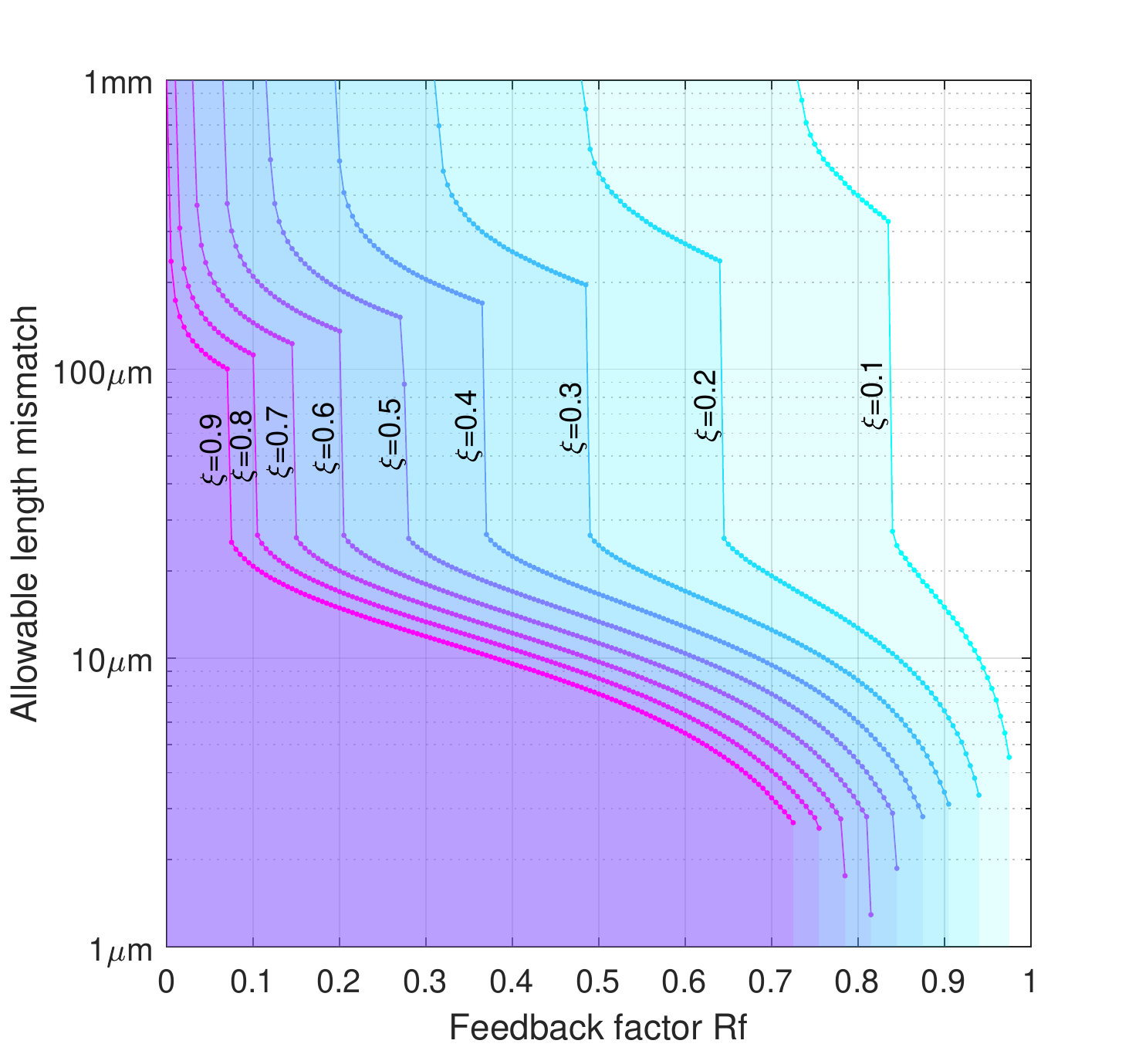}
	\caption{Allowable length mismatch to satisfy the stability condition about CFS with DOPO in a waveguide.}
	\label{fig:Stability_DOPO_waveguide}
\end{figure}
\begin{figure}[t]
	\centering
	\includegraphics[scale=0.6, clip]{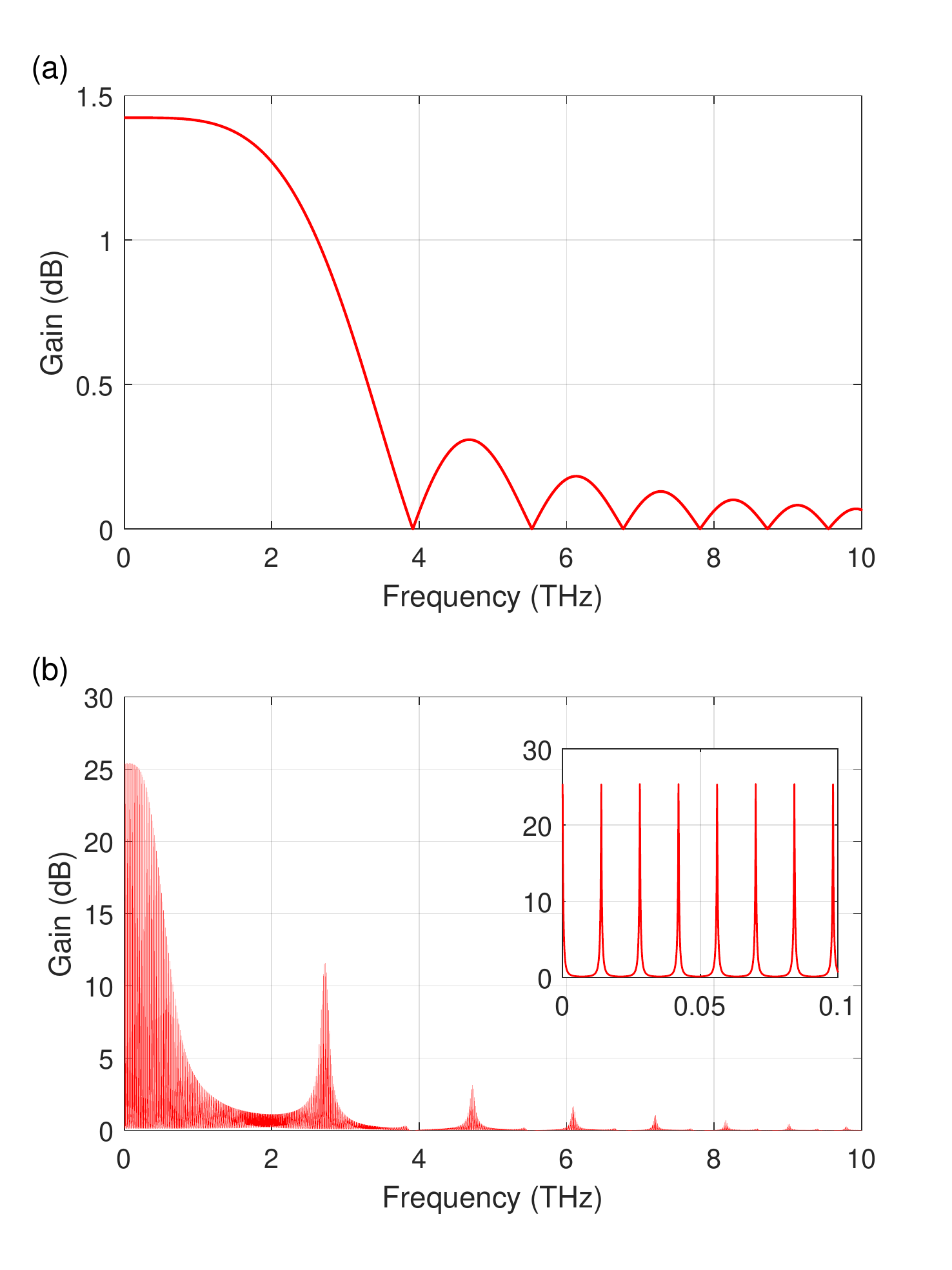}
	\caption{Amplifier gain spectrum with $\xi = 0.9$. (a) Single pass gain spectrum $20\log_{10} G^{(s)}_{\tilde x}(\omega)$. (b) DOPO gain spectrum $20\log_{10} G^{(o)}_{\tilde x}(\omega)$. The DOPO spectrum has resonance peaks whose interval is a FSR (inset). The envelope is determined by the parametric amplification gain spectrum and the wavelength dispersion in DOPO.}
	\label{fig:GainSpectrum}
\end{figure}

\figureref{Stability_DOPO_waveguide} shows the numerically calculated result of allowable length mismatch to satisfy the stability condition about CFS in a waveguide. Unlike the case of free space, we calculate the stability condition with the discrete values of normalized pump amplitude $\xi=0.1, \cdots, 0.9$. As in the free space case, the requirement regarding the fabrication precision can be read from this figure, e.g., if the fabrication error is $\pm 10~\mu$m, the CFS is stable in the region of ($R_{f}< 0.38$ and $\xi < 0.9$) or ($R_{f}< 0.94$ and $\xi < 0.1$).

Notably, the dependency of the allowable length mismatch on $R_{f}$ has discontinuous transitions at around 20-30~$\mu$m and 100-300~$\mu$m. This can be understood by looking at the DOPO gain spectrum. \figureref{GainSpectrum} shows the gain spectrums of the single pass gain $20\log_{10} G^{(s)}_{\tilde x}(\omega)$ and DOPO $20\log_{10} G^{(o)}_{\tilde x}(\omega)$ in the case of $\xi = 0.9$ [App.~\ref{sec:AppGainSpec}]. The wave vector mismatch $\Delta k_\omega $ determines the single pass gain spectrum as $\mathrm{sinc}^2(\Delta k_\omega l_c /2)$, $(\mathrm{sinc} (x) = (\sin x)/x)$, which also gives the envelope of the DOPO gain peaks if there is no wavelength dispersion. However, under the dispersion, since the resonance condition for the feedback loop of upper frequency differs from the lower one ($\exp[i\omega n_{\omega_c + \omega}l_o/c]=1$ for upper, $\exp[i\omega n_{\omega_c - \omega}l_o/c]=1$ for lower, see \figref{EqDdiagram} in App.~\ref{sec:TransferFunction}), the envelope of the DOPO gain peaks does not match the single pass gain spectrum. It means that the envelope also has another resonance condition $\exp [i\omega (n_{\omega_c + \omega}-n_{\omega_c - \omega})l_o/c]=1$. The first peak of the envelope has less than 1~THz bandwidth which is narrower than the bandwidth of the single pass gain of about 2.5~THz, which leads to relaxing the requirement for the stability compared with the case without dispersion at propagation. The next peak of the envelope around 2.7~THz is within the first comb of the single pass gain spectrum. If the peak affects the stability condition, the requirement for the stability will be severe. Note again that $\sqrt{R_f}$ times the DOPO gain is (roughly) the loop gain. In the case of $R_{f}<7\%$ which corresponds to the attenuation of more than about $-12$~dB, the resonance of the envelope around 2.7~THz is negligible, so the required precision is more than 100~$\mu$m determined by the first peak of the envelope. Thus, the discontinuous transition in \figref{Stability_DOPO_waveguide} shows the transition between these two situations. 

In the case of free space optics, if the fabrication error is less than 0.1~mm, the stability condition is almost satisfied in the whole area. On the other hand, in the case of a waveguide, the order is 10~$\mu$m, which is an order of magnitude smaller. This is because the 100~GHz optical BPF was inserted into the feedback loop in free space optics, while the phase-matching bandwidth is the same in both cases. Considering that the typical fabrication accuracy integrated optical devices is on the order of tens of nanometers, this constraint should not prevent the realization of an integrated CFS.

\section{Conclusion}
We investigate a new type of squeezer called a coherent feedback squeezer and show that the CFS has the several advantages over the existing squeezer, in particular, in terms of sensitivity and bandwidth. To conduct the analysis, we utilize frequency representation of an LTI quantum coherent feedback system. We derive input-output relations of the system, which are represented by the transfer functions, and the stability condition that the system does not oscillate. Our formulation enables precise analysis of the stability condition in terms of the loop gains, which are functions of frequency and incorporate experimental factors such as wavelength dispersion and phase-matching bandwidth.

As a demonstration of the feasibility study, we conduct the stability analysis of CFS based on both a free-space optics and waveguide devices. The result of the feasibility studies are interpreted as the tolerance to fabrication errors. This gives us the knowledge of what is the best design candidates as well as the physics insight of the system.

Although the feasibility studies in this paper were conducted on the optical devices, our formalism based on frequency representation is applicable also to different bosonic systems as well. The toolbox for theoretical analysis developed in this paper  provides useful methods for designing and developing feasible devices in quantum engineering.

\section*{Acknowledgments}
The authors would like to thank Matthew James and Naoki Yamamoto for their valuable comments.
This work was supported financially by the Australian Research Council Centres of Excellence scheme number CE170100012. M. L. was supported by the Australian Research Council Future Fellowship (FT180100055).
D.P.  was supported by the Australian Government Research Training Program Scholarship.
W. A. acknowledges financial support from the Japan Society for the Promotion of Science (JSPS).
T. T. was supported by the JSPS (16J09150).

\appendix
\section{Theoretical Formulations of Quantum Coherent Feedback System}
\label{sec:App_Theory}
\begin{figure}[b]
	\centering
	\includegraphics[scale=1.0, clip]{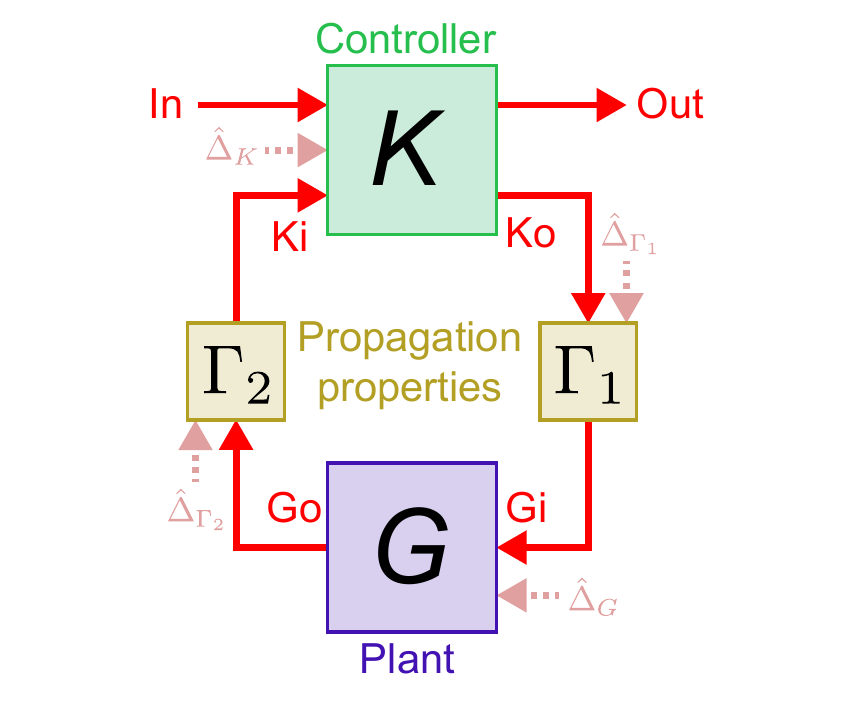}
	\caption{Block diagram of quantum coherent feedback system consisting of a linear quantum amplifier $G$ (plant), controller $K$, and filters $\Gamma_1$ and $\Gamma_2$ describing propagation properties. $\hat \Delta_G$, $\hat \Delta_K$, $\hat \Delta_{\Gamma_1}$, and $\hat \Delta_{\Gamma_2}$ are unwanted noise operators. The entire system is linear time-invariant, and all components except for a plant are passive.}
	\label{fig:System}
\end{figure}

In this Appendix, we will derive the full representation of a quantum coherent feedback system with a linear quantum amplifier. \figureref{System} shows the block diagram of our interest. We consider a quantum coherent feedback system with a plant that includes a quantum amplifier. This is a generalization of CFS shown in \figref{QOPamp}(a). We assume a general linear quantum amplifier, e.g., phase-sensitive and phase insensitive amplifiers. The derivations in this section assume realistic imperfections; unwanted noises at an amplifier and a controller, and losses and time-delays for propagations between an amplifier and a controller. The system is assumed to be a LTI system and composed of passive components except for the plant. The descriptions are based on general open linear quantum systems. They involve a broad range of classes written by bosonic annihilation and creation operators such as optical devices \cite{Bachor04,Furusawa11}, mechanical oscillators \cite{Gardiner04,Aspelmeyer14,Iwasawa13} and atomic ensembles \cite{Hammerer10,Yamamoto14atom,Yamamoto14}. Although we mainly focus on the optical descriptions, the following calculations are not limited to optics.

\subsection{System Representations}
\subsubsection{Plant}
Plant is defined as a single input single output (SISO) and LTI system whose input-output relation is represented in the frequency domain as
\begin{align}
\label{eq:Plant}
		\hat A_{Go}(\omega) = G_a(\omega) \hat A_{Gi} (\omega)+ G_c(\omega) \hat A^{\dagger}_{Gi} (-\omega ) + \hat \Delta_G (\omega).
	\end{align}
Here, the subscriptions are the indices of input $Gi$ and output $Go$ for the plant. $\hat \Delta_G$ is an operator describing unwanted noise from the environment which commutes with $\hat A_{Gi}(\omega ) $ and $\hat A^\dagger_{Gi} (-\omega ) $.
To preserve the commutation relations $[\hat A_k(\omega), \hat A_l^\dagger(\omega')] = \delta_{k,l} \delta(\omega-\omega')$ ($\delta_{k,l}$: Kronecker delta) and $[\hat A_k(\omega), \hat A_l(\omega')] = [\hat A_k^\dagger(\omega), \hat A_l^\dagger(\omega')] =0$ ($k,l \in \{Go,Gi\}$), the noise operator must satisfy the following equations:
\begin{align}
\notag
	&[\hat \Delta_G (\omega), \hat  \Delta_G^\dagger (\omega')] \\
	&\quad=[1-|G_a(\omega)|^2+|G_c(\omega)|^2]\delta (\omega-\omega'), \\
\notag
	&[\hat \Delta_G (\omega), \hat  \Delta_G (\omega')] \\
	&\quad=[G_a(-\omega)G_c(\omega) - G_a(\omega)G_c(-\omega)]\delta (\omega+\omega'), \\
\notag
	&[\hat \Delta_G^\dagger (\omega), \hat  \Delta_G^\dagger (\omega')] \\
	&\quad=[G^*_a(\omega)G^*_c(-\omega) - G^*_a(-\omega)G^*_c(\omega)]\delta (\omega+\omega').
\end{align}

\subsubsection{Controller}
Controller is defined as a two input two output and passive LTI system whose input-output relation can be represented in the frequency domain as 
	\begin{align}
	\label{eq:Controller}
		\begin{pmatrix}
			\hat A_{\mathrm{out}} (\omega) \\ \hat A_{Ko} (\omega)
		\end{pmatrix}
		&=
		\bm{K}(\omega )
		\begin{pmatrix}
			\hat A_{\mathrm{in}} (\omega) \\ \hat A_{Ki} (\omega)
		\end{pmatrix} + \hat{\bm{\Delta}}_K (\omega), \\
	\bm{K} (\omega )&\equiv
		\begin{pmatrix}
			K_{11}(\omega ) & K_{12}(\omega ) \\ K_{21}(\omega ) & K_{22}(\omega )
		\end{pmatrix}, \\
	\hat{\bm{\Delta}}_K (\omega) & \equiv
		\begin{pmatrix}
			\hat \Delta_{K_1}(\omega ) \\ \hat \Delta_{K_2}(\omega )
		\end{pmatrix}.
	\end{align}
The subscriptions are the indices of inputs and outputs for the controller, as shown in \figref{System}, where the indices of $in$ and $out$ means input and output modes for the entire system, respectively. $\hat \Delta_{K_1}(\omega )$ and $\hat \Delta_{K_2}(\omega )$ are noise operators commuting with $\hat A_{\mathrm{in}}(\omega )$ and $\hat A_{Ki}(\omega )$, which vanish in the ideal case.
$\bm{K} (\omega )$ and $\hat{\bm{\Delta}}_K(\omega )$ must satisfy the following conditions
\begin{align}
\notag
	&
	\begin{pmatrix}
	[\hat \Delta_{K_1} (\omega), \hat  \Delta_{K_1}^\dagger (\omega')]  & [\hat \Delta_{K_1} (\omega), \hat  \Delta_{K_2}^\dagger (\omega')] \\
	[\hat \Delta_{K_2} (\omega), \hat  \Delta_{K_1}^\dagger (\omega')]  & [\hat \Delta_{K_2} (\omega), \hat  \Delta_{K_2}^\dagger (\omega')] 
	\end{pmatrix}
	=\\
	&\qquad \left[
	\begin{pmatrix}
	1 & 0\\
	0 & 1
	\end{pmatrix}
	-	\bm{K}(\omega)\bm{K}^\dagger (\omega)
	\right]
	\delta (\omega -\omega'), \\
	&[\hat \Delta_{K_j} (\omega), \hat  \Delta_{K_{j'}} (\omega')]  = [\hat \Delta_{K_j}^\dagger (\omega), \hat  \Delta_{K_{j'}}^\dagger (\omega')]  = 0,
\end{align}
which are derived from the conditions to preserve commutation relations at the output.

\subsubsection{Other passive components}
Additional passive components, one before and one after the plant, are defined as SISO and passive LTI systems which input-output relations [\figref{System}] can be represented in the frequency domain as
	\begin{align}
	\label{eq:Propagation1}
		\hat A_{Gi} (\omega) &= \Gamma_1(\omega) \hat A_{Ko}(\omega) +\hat \Delta_{\Gamma_1} (\omega) , \\
	\label{eq:Propagation2}
		\hat A_{Ki} (\omega) &=  \Gamma_2(\omega) \hat A_{Go}(\omega) +\hat \Delta_{\Gamma_2} (\omega) ,
	\end{align}
where $\Gamma_1$ and $\Gamma_2$ are the transfer functions for the propagation, 
and $\hat \Delta_{\Gamma_1}$ and $\hat \Delta_{\Gamma_2}$ are noise terms for $\Gamma_1$ and $\Gamma_2$.
To satisfy the energy conservation law, the absolute values of the transfer function must be less than or equal to 1 ($|\Gamma (\omega ) |\le 1$).
From the preservation of commutation relations at the output, the noise terms must satisfy the following conditions
\begin{align}
	&[\hat \Delta_j (\omega), \hat  \Delta_j^\dagger (\omega')]=\left[1-|\Gamma_j (\omega)|^2\right]\delta (\omega-\omega'), \\
	&[\hat \Delta_j (\omega), \hat  \Delta_j (\omega')] = [\hat \Delta_j^\dagger (\omega), \hat  \Delta_j^\dagger (\omega')] = 0,  \quad (j=\Gamma_1,\Gamma_2).
\end{align}

\subsection{Transfer Function of the Entire System}
\label{sec:TransferFunction}
\begin{figure}[t]
	\centering
	\includegraphics[scale=1.0, clip]{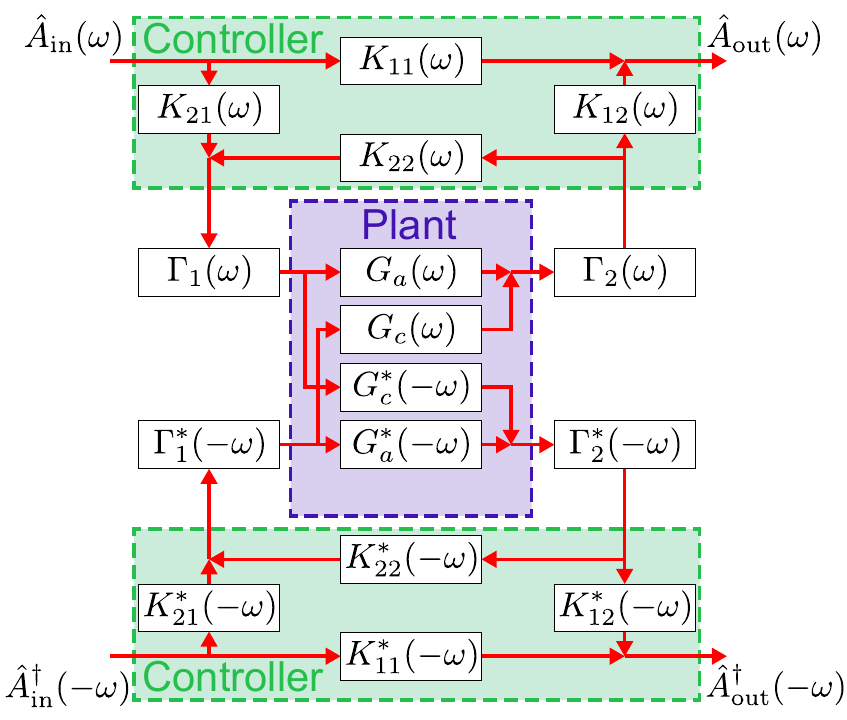}
	\caption{The equivalent block diagram in the frequency domain. Unwanted noise operators are omitted for simplicity.}
	\label{fig:EqDdiagram}
\end{figure}

By using Eqs.~\eqref{eq:Plant},\eqref{eq:Controller},\eqref{eq:Propagation1}, and \eqref{eq:Propagation2}, we derive the input-output relation for the entire system as
	\begin{align}
	\notag
		&\hat A_{\mathrm{out}} (\omega ) = 
		   G^{(\mathrm{fb})}_{a}(\omega ) \hat A_{\mathrm{in}} (\omega ) +G^{(\mathrm{fb})}_{c}(\omega ) \hat A^\dagger_{\mathrm{in}} (-\omega ) \\
	\label{eq:InputOutput_A}
\notag
		&+ \sum _{j}\left[G^{(\mathrm{fb})}_{ja}(\omega ) \hat \Delta_j (\omega ) +G^{(\mathrm{fb})}_{jc}(\omega ) \hat \Delta^\dagger_j (-\omega ) \right]. \\
	\end{align}
where $j = G, \Gamma_1,\Gamma_2,K_1,K_2$, and transfer functions are
	\begin{align}
	\notag
		&G^{(\mathrm{fb})}_{a}(\omega ) =\tfrac{1}{K_{22}(\omega )}\left[ \det K(\omega ) +\tfrac{1-\Lambda_a^*(-\omega )}{\Lambda_D(\omega )}K_{12}(\omega)K_{21}(\omega)\right],\\
	\notag
		&G^{(\mathrm{fb})}_{c}(\omega ) =\tfrac{\Lambda_c (\omega)}{\Lambda_D(\omega)}\tfrac{K_{12}(\omega)K^*_{21}(-\omega)}{K_{22}(\omega)}\tfrac{\Gamma^*_1(-\omega)}{\Gamma_1(\omega)},\\
	\notag
		&G^{(\mathrm{fb})}_{\Gamma_1a}(\omega ) =\left[\tfrac{1-\Lambda_a^*(-\omega )}{\Lambda_D(\omega )}-1\right]\tfrac{K_{12}(\omega)}{K_{22}(\omega)} \tfrac{1}{\Gamma_1(\omega)} ,\\
	\notag
		&G^{(\mathrm{fb})}_{\Gamma_1c}(\omega ) =\tfrac{\Lambda_c(\omega )}{\Lambda_D(\omega )}\tfrac{K_{12}(\omega)}{K_{22}(\omega)} \tfrac{1}{\Gamma_1(\omega)} ,\\
	\notag
		&G^{(\mathrm{fb})}_{K_2a}(\omega ) =\left[\tfrac{1-\Lambda_a^*(-\omega )}{\Lambda_D(\omega )}-1\right]\tfrac{K_{12}(\omega)}{K_{22}(\omega)},\\ 
	\notag
		&G^{(\mathrm{fb})}_{K_2c}(\omega ) =\tfrac{\Lambda_c (\omega)}{\Lambda_D(\omega)}\tfrac{K_{12}(\omega)}{K_{22}(\omega)} \tfrac{\Gamma^*_1(-\omega)}{\Gamma_1(\omega)},\\ 
	\notag
		&G^{(\mathrm{fb})}_{\Gamma_2a}(\omega ) =\tfrac{1-\Lambda_a^*(-\omega )}{\Lambda_D(\omega )}K_{12}(\omega),\\
	\notag
		&G^{(\mathrm{fb})}_{\Gamma_2c}(\omega ) = \tfrac{\Lambda_c (\omega)}{\Lambda_D(\omega)}\tfrac{K_{12}(\omega)K^*_{22}(-\omega)}{K_{22}(\omega)}\tfrac{\Gamma^*_1(-\omega)}{\Gamma_1(\omega)},\\
	\notag
		&G^{(\mathrm{fb})}_{Ga}(\omega ) =
\tfrac{1-\Lambda_a^*(-\omega )}{\Lambda_D(\omega )}K_{12}(\omega) \Gamma_2(\omega),\\
	\notag
		&G^{(\mathrm{fb})}_{Gc}(\omega ) =
 \tfrac{\Lambda_c (\omega)}{\Lambda_D(\omega)}\tfrac{K_{12}(\omega)K^*_{22}(-\omega)}{K_{22}(\omega)}\tfrac{\Gamma^*_1(-\omega)\Gamma^*_2(-\omega)}{\Gamma_1(\omega)}, \\
	\notag
		&G^{(\mathrm{fb})}_{K_1a}(\omega ) =1,\\
\label{eq:TransferFunctionFB}
		&G^{(\mathrm{fb})}_{K_1c}(\omega ) =0.
	\end{align}
Here, we define $\Lambda_a$, $\Lambda_c$,  and $\Lambda_D$ as
	\begin{align}
	\label{eq:Ka}
		\Lambda_a (\omega) &\equiv  K_{22}(\omega )\Gamma_1(\omega) \Gamma_2(\omega)G_a(\omega), \\
	\label{eq:Kc}
		\Lambda_c (\omega) &\equiv  K_{22}(\omega )\Gamma_1(\omega) \Gamma_2(\omega)G_c(\omega), \\
\notag
		\Lambda_D(\omega ) &\equiv [1-\Lambda_a(\omega )][1-\Lambda^*_a(-\omega )]  \\
	\label{eq:KD}
		& \qquad - \Lambda_c(\omega )\Lambda^*_c(-\omega ) .
	\end{align}
$\Lambda_a(\omega)$ and $\Lambda^*_a(-\omega)$ correspond to the loop gains for $\hat A(\omega)$ and $\hat A^\dagger (-\omega)$, respectively. On the other hand, $\Lambda_c(\omega)\Lambda^*_c(-\omega)$ also corresponds to another loop gain [\figref{EqDdiagram}].
We also define the transfer function matrix as
\begin{align}
	\bm{\Lambda}_{a,c}(\omega ) \equiv 
	\begin{pmatrix}
		\Lambda_a(\omega) & \Lambda_c(\omega )  \\
		\Lambda_c^* (-\omega) & \Lambda_a^* (-\omega ).
	\end{pmatrix}
\end{align}
By using this matrix, $\Lambda_D(\omega )$ is expressed as
\begin{align}
\label{eq:KDTraceDet}
	\Lambda_D (\omega ) = 1-\mathrm{Tr}[\bm{\Lambda}_{a,c}(\omega )]  + \det [\bm{\Lambda}_{a,c}(\omega )].
\end{align}

\subsection{Representation by Quadratures}
The input-output relation in Eq.~\eqref{eq:InputOutput_A} is represented by a matrix form as
\begin{align}
\label{eq:TransferFunctionA}
\bm{A}_{\mathrm{out}}(\omega )	
	= \bm{G}^{(\mathrm{fb})}(\omega ) \bm{A}_{\mathrm{in}}(\omega )	,
\end{align}
where, $\bm{G}^{(\mathrm{fb})}(\omega )$ is a corresponding transfer function matrix, and
\begin{align}
	 \bm{A}_{\mathrm{out}}(\omega )	\equiv (&\hat A_{\mathrm{out}} (\omega ), \hat A^\dagger_{\mathrm{out}} (-\omega ))^T, \\
\notag
\bm{A}_{\mathrm{in}}(\omega )	 \equiv  (&
\hat A_{\mathrm{in}} (\omega ),\hat A^\dagger_{\mathrm{in}} (-\omega ),
\hat \Delta_{G} (\omega ),\hat \Delta^\dagger_{G} (-\omega ), \\
\notag
&\hat \Delta_{\Gamma_1} (\omega ),\hat \Delta^\dagger_{\Gamma_1} (-\omega ),
\hat \Delta_{\Gamma_2} (\omega ),\hat \Delta^\dagger_{\Gamma_2} (-\omega ),\\
&\hat \Delta_{K_1} (\omega ),\hat \Delta^\dagger_{K_1} (-\omega ),
\hat \Delta_{K_2} (\omega ),\hat \Delta^\dagger_{K_2} (-\omega )
)^T.
\end{align}
The relation between annihilation-creation operators and quadratures is expressed as 
\begin{align}
	\bm{Q}_{\mathrm{out}}(\omega )=
	\bm{J}
	\bm{A}_{\mathrm{out}}(\omega ), 
\end{align}
where
\begin{align}
	\bm{Q}_{\mathrm{out}}(\omega )
	 \equiv 
	\begin{pmatrix}
		\hat X_{\mathrm{out}} (\omega )  \\ \hat P_{\mathrm{out}} (\omega )
	\end{pmatrix},
	\quad
	\bm{J} \equiv \frac{1}{2}
	\begin{pmatrix}
		1 & 1 \\ - i  & i
	\end{pmatrix}.
\end{align}
Equation \eqref{eq:TransferFunctionA} is rewritten in the form of quadratures as
\begin{align}
	\bm{Q}_{\mathrm{out}}(\omega ) = \bm{G}_q^{(\mathrm{fb})}(\omega ) \bm{Q}_{\mathrm{in}}(\omega ),
\end{align}
where
\begin{align}
	\label{eq:TransferFunctionMatrixQ}
	\bm{G}_q^{(\mathrm{fb})}(\omega ) \equiv \bm{J} \bm{G}^{(\mathrm{fb})}(\omega ) \diag [\bm{J}^{-1},\cdots, \bm{J}^{-1}],
\end{align} and
\begin{align}
\notag
\bm{Q}_{\mathrm{in}}(\omega )	 \equiv  (&
\hat X_{\mathrm{in}} (\omega ),\hat P_{\mathrm{in}} (\omega ),
\hat \Delta^{(X)}_{G} (\omega ),\hat \Delta^{(P)}_{G} (\omega ), \\
\notag
&\hat \Delta^{(X)}_{\Gamma_1} (\omega ),\hat \Delta^{(P)}_{\Gamma_1} (\omega ),
\hat \Delta^{(X)}_{\Gamma_2} (\omega ),\hat \Delta^{(P)}_{\Gamma_2} (\omega ),\\
&\hat \Delta^{(X)}_{K_1} (\omega ),\hat \Delta^{(P)}_{K_1} (\omega ),
\hat \Delta^{(X)}_{K_2} (\omega ),\hat \Delta^{(P)}_{K_2} (\omega )
)^T.
\end{align}Here, we use the noise operators in the quadrature form,
\begin{align}
	\begin{pmatrix}
		\hat \Delta^{(X)}_j (\omega) \\ \hat \Delta^{(P)}_j (\omega)
	\end{pmatrix}
	\equiv 
	\bm{J}
	\begin{pmatrix}
		\hat \Delta_j (\omega) \\ \hat \Delta^\dagger_j (-\omega)
	\end{pmatrix},
	\quad 
	(j= G,\Gamma_1,\Gamma_2,K_1,K_2).
\end{align}
The transfer function from input to output at the plant is expressed as
\begin{align}
	\label{eq:Gmatrix}
	\begin{pmatrix}
		G_x(\omega)  & G_{p\to x} (\omega ) \\
		G_{x\to p} (\omega )  & G_p(\omega)
	\end{pmatrix}	
	\equiv
	\bm{J}
	\begin{pmatrix}
		G_a(\omega)  & G_c (\omega ) \\
		G_c^* (-\omega )  & G_a^*(-\omega)
	\end{pmatrix}		
	\bm{J}^{-1}.
\end{align}
In the same manner, we define the loop gain matrix for the quadratures as
\begin{align}
\label{eq:LoopgainsXP}
\bm{\Lambda}_{x,p} (\omega ) \equiv
	\begin{pmatrix}
		\Lambda_x(\omega)  & \Lambda_{p\to x} (\omega ) \\
		\Lambda_{x\to p} (\omega )  & \Lambda_p(\omega)
	\end{pmatrix}	
	=
	\bm{J}
	\bm{\Lambda}_{a,c}(\omega)
	\bm{J}^{-1}.
\end{align}
Since this transformation preserves both trace and determinant of the transfer function matrix, we can express $\Lambda_D$ by loop gains for the quadratures in the same manner as Eq.~\eqref{eq:KDTraceDet}:
\begin{align}
\notag
		\Lambda_D(\omega ) &= [1-\Lambda_x(\omega )][1-\Lambda_p(\omega )]  
	 - \Lambda_{x\to p}(\omega )\Lambda_{p\to x}(\omega ) \\
\label{eq:KDquadratures}
	&=1-\mathrm{Tr}[\bm{\Lambda}_{x,p}(\omega )]  + \det [\bm{\Lambda}_{x,p}(\omega )].
\end{align}

\subsection{Stability}
\label{sec:Stability}
As we mentioned in the main text, in order for a quantum LTI system to be stable, all of the roots of the characteristic equation 
	\begin{align}
	\label{eq:characteristicEq}
		\Lambda_D(\omega = is) = 0,
	\end{align}
must have negative real parts \cite{Yamamoto16,Yokotera19,Yoshimura19,Shimazu19}. Note again, that we may omit $\hat{\bm{\Delta}}_K (\omega )$, $\hat \Delta_{\Gamma_1} (\omega )$, $\hat \Delta_{\Gamma_2} (\omega )$ in certain systems. These noise terms do not affect stability analysis; that is, the characteristic equation $\Lambda_D$ does not include any information about the noise operators. Of course, the loss represented by $\Gamma_1$ and $\Gamma_2$ appears in the characteristic equation. The significant fact is that the characteristic equation consists of only the loop gains ($\Lambda_a(\omega)$, $\Lambda_a^*(-\omega)$, $\Lambda_c(\omega)\Lambda_c^*(-\omega))$ or ($\Lambda_x(\omega)$, $\Lambda_p(\omega)$, $\Lambda_{x\to p}(\omega)\Lambda_{p\to x}(\omega)$); It is enough to investigate the loop gains for the stability analysis.

In an LTI system, the ensemble average of quantum fluctuation does not depend on time, which allows measuring these transfer functions directly by injecting probes like a coherent beam which has an expectation value substantially larger than quantum fluctuations as long as the system keeps linearity. That is, we assume that the transfer function obtained from the classical measurement is the same as the transfer function regarding the expectation values of the input-output for a quantum system \cite{Lobino08}. Under this assumption,  there is no need to measure the quantum responses to the quantum signals. In the main text, we conduct the stability analysis for our optical CFS by means of the loop gains derived from the information such as an optical bandpass filter and wavelength dispersion.

\subsection{Symmetry around the Carrier Frequency}
\label{sec:AppSymmetry}
Equations~\eqref{eq:InputOutput_A} and \eqref{eq:TransferFunctionMatrixQ} are the transfer functions in general representation. As seen in many of the optical systems, the transfer function has typically the symmetry around the carrier frequency, which makes the expression simpler. We define the symmetry of the system as follows. 

\mbox{}

The frequency-symmetric system is defined as the system in which transfer functions $O(\omega ) $ have the following symmetry around the carrier frequency:
\begin{align}
\label{eq:DefSymmetry}
	O^* (-\omega ) = O (\omega ), \quad (O=K,\Gamma_{1},\Gamma_{2},G_a,G_c).
\end{align}

\mbox{}

\begin{figure}[t]
	\centering
	\includegraphics[scale=1.0, clip]{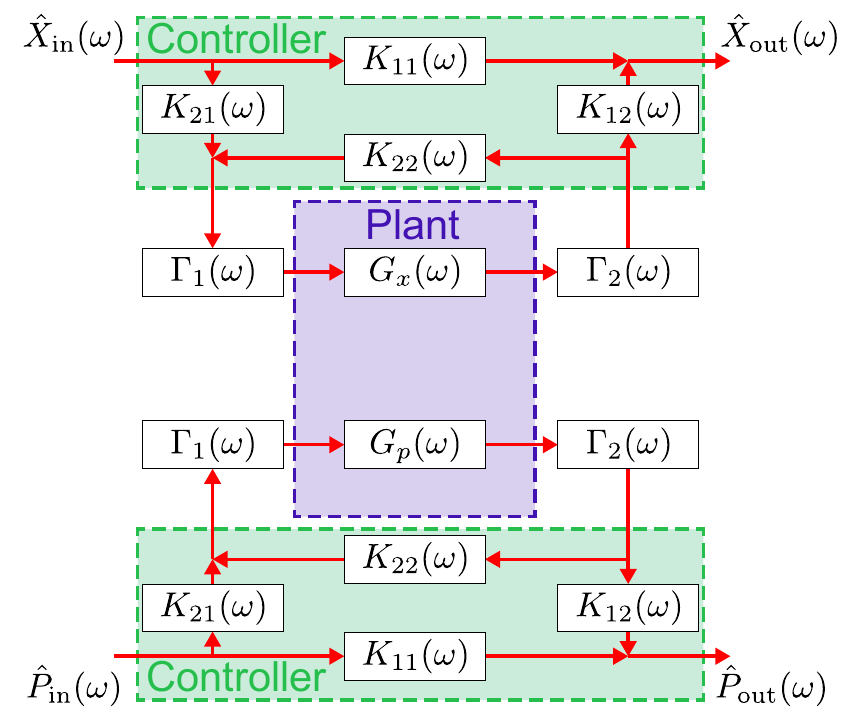}
	\caption{The equivalent block diagram under the symmetry around the carrier frequency. Unwanted noise operators are omitted for simplicity. The plant is considered as a SISO system for each quadrature.}
	\label{fig:EqDdiagram2}
\end{figure}

In the following, we assume the symmetry around the carrier frequency for all of the frequency functions. The symmetry simplifies the transfer function of the plant for the quadratures:
\begin{align}
		G_{x}(\omega ) &= G_{a}(\omega ) +G_{c}(\omega ), \\
		G_{p}(\omega ) &= G_{a}(\omega ) -G_{c}(\omega ), \\
		G_{x\to p}(\omega ) & = G_{p\to x}(\omega )= 0.
\end{align}
Under the symmetry around the carrier frequency, the transfer function matrix [Eq.~\eqref{eq:Gmatrix}] is diagonalized. The situation is the same in the other components. \figureref{EqDdiagram2} shows the block diagram of the system with the symmetry around the carrier frequency. All of the transformations are closed in each $\hat X$ or $\hat P$ quadratures.
As a result, the transfer function of the entire system [Eq.~\eqref{eq:InputOutput_A}] is composed of only $\hat X$ or $\hat P$:
	\begin{align}
	\label{eq:TransferfunctionXout}
		&\hat X_{\mathrm{out}} (\omega ) = 
		G^{(\mathrm{fb})}_{x}(\omega ) \hat X_{\mathrm{in}} (\omega ) 
		+ \sum _{j}G^{(\mathrm{fb})}_{jx}(\omega ) \hat{\Delta}_j^{(X)} (\omega), \\
		&\hat P_{\mathrm{out}} (\omega ) = 
		G^{(\mathrm{fb})}_{p}(\omega ) \hat P_{\mathrm{in}} (\omega ) 
		+ \sum _{j}G^{(\mathrm{fb})}_{jp}(\omega ) \hat{\Delta}_j^{(P)}  (\omega),
	\end{align}
where $j  = G, \Gamma_1,\Gamma_2, K_1,K_2$, and 
\begin{align}
		G^{(\mathrm{fb})}_{x}(\omega ) &= G^{(\mathrm{fb})}_{a}(\omega ) +G^{(\mathrm{fb})}_{c}(\omega ), \\
		G^{(\mathrm{fb})}_{jx}(\omega ) &= G^{(\mathrm{fb})}_{ja}(\omega ) +G^{(\mathrm{fb})}_{jc}(\omega ), \\
		G^{(\mathrm{fb})}_{p}(\omega ) &= G^{(\mathrm{fb})}_{a}(\omega ) -G^{(\mathrm{fb})}_{c}(\omega ), \\
		G^{(\mathrm{fb})}_{jx}(\omega ) &= G^{(\mathrm{fb})}_{ja}(\omega ) -G^{(\mathrm{fb})}_{jc}(\omega ).
\end{align}
These transfer functions are written explicitly as follows:
	\begin{align}
	\notag
		G^{(\mathrm{fb})}_q (\omega ) &= K_{11}(\omega ) +\tfrac{K_{12}(\omega) K_{21}(\omega ) }{K_{22}(\omega ) }\tfrac{\Lambda_q(\omega )}{1-\Lambda_q(\omega )},\\
	\notag
		G^{(\mathrm{fb})}_{\Gamma_1q}(\omega ) &= \tfrac{1}{K_{21}(\omega ) \Gamma_1(\omega ) }\left[G^{(\mathrm{fb})}_q (\omega ) - K_{11}(\omega)  \right] \\
	\notag
	& = \tfrac{K_{12}(\omega)}{K_{22}(\omega) \Gamma_1 (\omega ) }\tfrac{\Lambda_q(\omega )}{1-\Lambda_q(\omega )},\\
	\notag
		G^{(\mathrm{fb})}_{K_2q}(\omega )  &= \Gamma_1 (\omega ) G^{(\mathrm{fb})}_{\Gamma_1q}(\omega ) \\
	\notag
	&=\tfrac{K_{12}(\omega)}{K_{22}(\omega)  }\tfrac{\Lambda_q(\omega )}{1-\Lambda_q(\omega )},\\
	\notag
		G^{(\mathrm{fb})}_{\Gamma_2q}(\omega ) &= \tfrac{K_{22}(\omega) }{K_{21}(\omega )} \left[ G^{(\mathrm{fb})}_q (\omega )- \tfrac{\det K}{K_{22}(\omega ) }\right]\\
	\notag
		&=K_{12}(\omega ) \tfrac{1}{1-\Lambda_q(\omega )},\\
	\notag
		G^{(\mathrm{fb})}_{Gq}(\omega )  &= \Gamma_2 (\omega ) G^{(\mathrm{fb})}_{\Gamma_2q}(\omega ) \\
	\notag
	&=K_{12}(\omega )\Gamma_2 (\omega)  \tfrac{1}{1-\Lambda_q(\omega )}, \\
\label{eq:TransferfunctionsQout}
	G^{(\mathrm{fb})}_{K_1q}(\omega )  &= 1,
	\end{align}
where $q=x,p$.

The loop gain matrix for the quadratures is also diagonalized:
\begin{align}
		\Lambda_{x}(\omega ) &= \Lambda_a(\omega ) +\Lambda_c(\omega ), \\
		\Lambda_{p}(\omega ) &= \Lambda_a(\omega ) -\Lambda_c(\omega ), \\
		\Lambda_{x\to p}(\omega ) & = \Lambda_{p\to x}(\omega )= 0.
\end{align}
So, the characteristic function [Eq.~\eqref{eq:KDquadratures}] can be factorized as
	\begin{align}
	\label{eq:Dsymmetry}
		\Lambda_D(\omega ) 
		& =\left[1-\Lambda_x(\omega )\right]\left[1-\Lambda_p(\omega )\right],
	\end{align}
where
	\begin{align}
		\Lambda_x (\omega) &=  K_{22}(\omega )\Gamma_1(\omega) \Gamma_2(\omega)G_x(\omega), \\
		\Lambda_p (\omega) &=  K_{22}(\omega )\Gamma_1(\omega) \Gamma_2(\omega)G_p(\omega).
	\end{align}
$\Lambda_x (\omega) $ and $\Lambda_p (\omega) $ correspond to the loop gains for $\hat X(\omega) $ and $\hat P(\omega) $ quadratures. In our quantum feedback system for a SISO plant, the stability condition is given by loop gains for each quadrature $\hat X(\omega) $ and $\hat P(\omega) $ under the assumption of the symmetry around the carrier frequency.

We assume this symmetry around the carrier frequency in Sec.~\ref{sec:CFSfmain} and App.~\ref{sec:CFSf} for the sake of simplicity. On the other hand, we do not assume the symmetry in order to consider the wavelength dispersion in Sec.~\ref{sec:CFSwmain} and App.~\ref{sec:CFSw}.

\section{Derivation of Input-Output Relation as a Function of Frequency in Coherent Feedback Squeezer}
\label{sec:App_Frequency}
We have derived equations for the ``general coherent feedback system'' in App.~\ref{sec:App_Theory}. Here we will explain the coherent feedback ``squeezer''.

We consider a DOPO as a plant in our coherent feedback squeezer. The second-order nonlinear conversion for the intra-cavity field $\hat A$ is described  by quadratic Hamiltonian $\hat{\mathcal{H}}_{\mathrm{DOPO}} = i \epsilon/4 [(\hat A^\dagger)^2 - \hat A^2]$, where the pump amplitude is $\epsilon \in \mathbb{R}$. The Langevin equation and the constraint between the intra-cavity field and ouput fields are described as \cite{Collett84}
\begin{align}
\notag
	\frac{\dd \hat A}{\dd t} &= -2i[\hat A , \hat{\mathcal{H}}_{\mathrm{DOPO}}]  - (\gamma+i \omega _d) \hat A  \\
\label{eq:Langevin1}
	& \qquad + \sqrt{\gamma_{T_o}} \hat A_{Gi} +  \sqrt{\gamma_{L_o}}\hat A_{\Delta G}, \\
\label{eq:Langevin2}
	\hat A_{Go} &= \sqrt{\gamma_T}\hat A -\hat A _{Gi}.
\end{align}
Here, $\hat A_{Gi}$, $\hat A_{Go}$, and $\hat A_{\Delta G}$ are annihilation operators of an input, an output, and a vacuum, respectively, where the argument $t$ is omitted for simplicity. The decay rates are given by the experimental parameters as $\gamma_{T_o} = cT_o/l_o$, $\gamma_{L_o} = cL_o/l_o$, and $\gamma = (\gamma_{T_o}+\gamma_{L_o})/2$, where $c$ is the speed of light in vacuum, while $\omega_d$, $l_o$, $T_o$, and $L_o$ are the detuning from resonance frequency, the cavity round trip length, the energy transmissivity (output coupler of DOPO), and the intra-cavity energy loss, respectively. Under the assumption of zero detuning $\omega_d = 0$, the transfer functions in Eq.~\eqref{eq:Plant} are written as \cite{Collett84}
\begin{align}
	\label{eq:PSATransferFunction1}
	G_a^{(\mathrm{DOPO})} (\omega )  & = \frac{(\gamma_{T_o}/2)^2-(\gamma_{L_o}/2-i\omega)^2+\epsilon^2 }{(\gamma-i\omega)^2 -\epsilon ^2},\\
	\label{eq:PSATransferFunction2}
	G_c^{(\mathrm{DOPO})} (\omega )  & = \frac{\epsilon \gamma_{T_o}}{(\gamma-i\omega)^2 -\epsilon ^2},\\
\notag
	\hat{\Delta}_G^{(\mathrm{DOPO})} (\omega )  & = \frac{\sqrt{\gamma_{T_o}\gamma_{L_o}}(\gamma-i\omega)}{(\gamma-i\omega)^2 -\epsilon ^2}
	 \hat A_{\Delta G} (\omega ) \\
	\label{eq:PSATransferFunction3}
	&\quad + \frac{\epsilon \sqrt{\gamma_{T_o}\gamma_{L_o}} }{(\gamma-i\omega )^2 -\epsilon ^2}
	\hat A^\dagger _{\Delta G} (-\omega ).
\end{align}
DOPO itself must satisfy the stability condition; the root of the characteristic equation $(\gamma + s)^2 - \epsilon^2 = 0 $ must lie on the left side of a complex plane to be stable. Thus, $\xi \equiv \epsilon /\gamma <1$, where we define the normalized pump amplitude $ \xi  \in [0,1), (\epsilon \ge 0) $. 

We also consider the transfer function of the controller [Eq.~\eqref{eq:Controller}] as
\begin{align}
	\bm{K}(\omega ) =
		\begin{pmatrix}
			\sqrt{R_f} & \sqrt{1-R_f} \\ \sqrt{1-R_f} & -\sqrt{R_f}
		\end{pmatrix},
	\quad \hat{\bm{\Delta}}_K (\omega) =\bm{0},
\end{align}
and the transfer functions regarding propagations [Eqs.~\eqref{eq:Propagation1}\eqref{eq:Propagation2}] as
	\begin{align}
		\Gamma_1(\omega) &= \sqrt{1-L_1}\ee^{i\omega \tau_1} , \quad \hat \Delta_{\Gamma_1} (\omega)  = \sqrt{L_1}\hat A_{\Delta{\Gamma_1}} (\omega) , \\
		\Gamma_2(\omega) &= \sqrt{1-L_2}\ee^{i\omega \tau_2} , \quad \hat \Delta_{\Gamma_2} (\omega)  = \sqrt{L_2}\hat A_{\Delta{\Gamma_2}} (\omega ),
	\end{align}
where $L_1$ $(L_2)$ and $\tau_1$ $(\tau_2)$ are the optical loss and the propagation time at one before a plant with the length $l_1$ (one after a plant with the length $l_2$), respectively. Here, we assume the symmetry around the carrier frequency [Eq.~\eqref{eq:DefSymmetry}] in all of the system. Thus, the transfer functions of the entire system can be simplified in the form of quadratures.

We use the loop gains for quadratures as
\begin{align}
\label{eq:KxCFS}
	\Lambda_x^{(\mathrm{CFS})}(\omega ) &= -\sqrt{R_{f}(1-L_{f})}\ee^{i\omega\tau_{f}} G^{(\mathrm{DOPO})}_x(\omega ), \\
	\Lambda_p^{(\mathrm{CFS})}(\omega ) &= -\sqrt{R_{f}(1-L_{f})}\ee^{i\omega\tau_{f}} G^{(\mathrm{DOPO})}_p(\omega ),
\end{align}
where
\begin{align}
\label{eq:GxPSA}
	G^{(\mathrm{DOPO})}_x(\omega ) &= \frac{(\gamma_{T_o} - \gamma_{L_o})/(2 \gamma ) + i\omega/\gamma +\xi}{1- i\omega/\gamma -\xi}, \\
	G^{(\mathrm{DOPO})}_p(\omega ) &= \frac{(\gamma_{T_o} - \gamma_{L_o})/(2 \gamma) + i\omega/\gamma -\xi}{1- i\omega/\gamma +\xi}.
\end{align}
Here, we redefine $L_{f} = 1-(1-L_1)(1-L_2)$, $\tau_{f}= \tau_1 + \tau_2 = l_1/c + l_2/c = l_{f}/c$, where $L_{f}$, $\tau_{f}$, and $l_{f}$ are the total feedback loop loss, time, and length, respectively.

The transfer function from input to output is calculated from Eqs.~\eqref{eq:TransferfunctionXout}-\eqref{eq:TransferfunctionsQout} as 
\begin{align}
\notag
	&\hat X_{\mathrm{out}}(\omega ) =
		\frac{\sqrt{R_{f}}-\Lambda_x^{(\mathrm{CFS})}(\omega)/\sqrt{R_{f}}}{1-\Lambda_x^{(\mathrm{CFS})}(\omega)}\hat X_{\mathrm{in}}(\omega)  \\
\notag
 & +\frac{\sqrt{(1-R_{f})(1-L_{2})}\ee^{i\omega\tau_2}}{1-\Lambda_x^{(\mathrm{CFS})}(\omega)}G_{\Delta x}^{(\mathrm{DOPO})}(\omega ) \hat X_{\Delta G}(\omega)\\
\notag & +\frac{\sqrt{(1-R_{f})L_{1}(1-L_2)}\ee^{i\omega\tau_2}}{1-\Lambda_x^{(\mathrm{CFS})}(\omega)}G_x^{(\mathrm{DOPO})}(\omega)\hat X_{\Delta \Gamma_1}(\omega) \\
\label{eq:CFSOutput}
 & +\frac{\sqrt{(1-R_{f})L_{2}}}{1-\Lambda_x^{(\mathrm{CFS})}(\omega)}\hat X_{\Delta \Gamma_2}(\omega), 
\end{align}
where
\begin{align}
	G_{\Delta x}^{(\mathrm{DOPO})} &= \frac{\sqrt{\gamma_{T_o}\gamma_{L_o}}/\gamma}{1-i\omega/\gamma  - \xi}, \\
\label{eq:GdeltapPSA}
	G_{\Delta p}^{(\mathrm{DOPO})} &= \frac{\sqrt{\gamma_{T_o}\gamma_{L_o}}/\gamma}{1-i\omega/\gamma  + \xi}, \\
	\begin{pmatrix}
		\hat X_{\Delta j}(\omega) \\
		\hat P_{\Delta j}(\omega)
	\end{pmatrix}
	&=\bm{J}
	\begin{pmatrix}
		\hat A_{\Delta j}(\omega) \\
		\hat A^\dagger_{\Delta j}(-\omega)
	\end{pmatrix}.
\end{align}
In the same manner, we can get the input-output relation of $\hat P$ quadrature by replacing ($\hat X$, $\Lambda_x^{(\mathrm{CFS})}$, $G_x^{(\mathrm{DOPO})}$, $G_{\Delta x}^{(\mathrm{DOPO})}$) with ($\hat P$, $\Lambda_p^{(\mathrm{CFS})}$, $G_p^{(\mathrm{DOPO})}$, $G_{\Delta p}^{(\mathrm{DOPO})}$).

\section{Loop Gains and Stability Conditions in Realistic Systems}
\label{sec:LoopGains}
\begin{figure}[b]
	\centering
	\includegraphics[scale=1.0, clip]{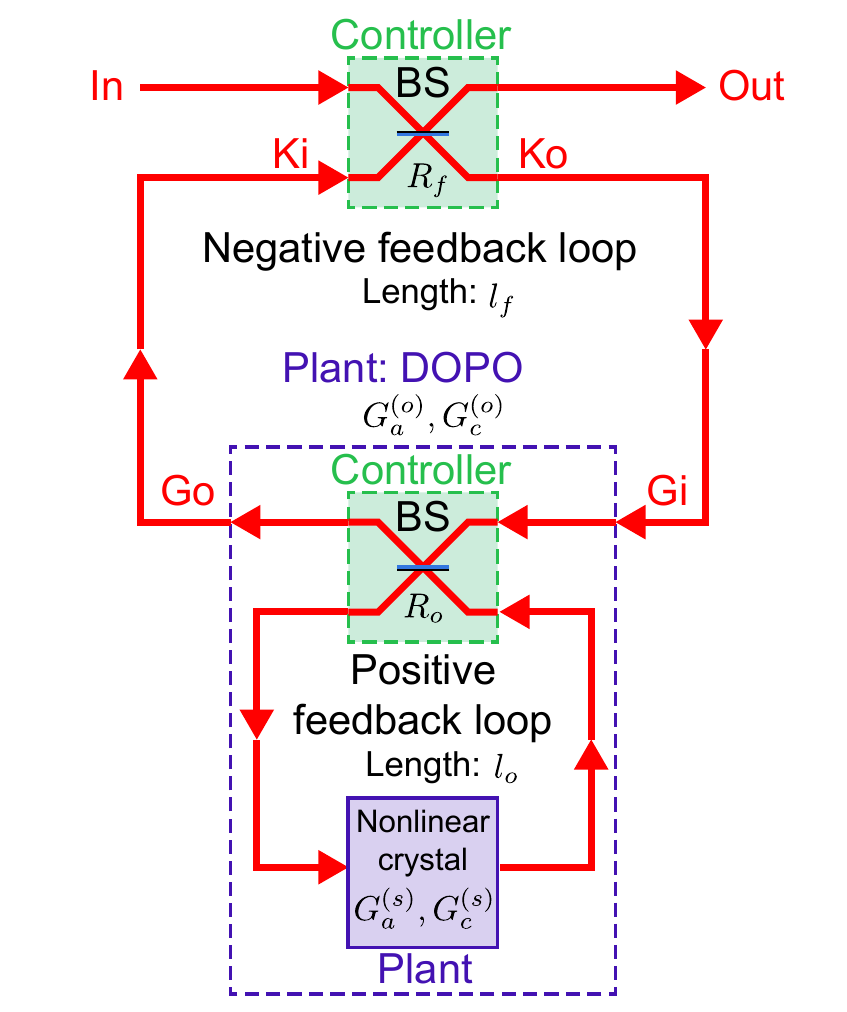}
	\caption{The abstract diagram of CFS involving DOPO. CFS is a quantum coherent negative-feedback system, while DOPO is also one of coherent feedback system with positive feedback polarity.}
	\label{fig:OPOandCFS}
\end{figure}

In App.~\ref{sec:App_Frequency}, we derived the general description of a CFS as a function of frequency from the quantum Langevin equation which is approximated to consider only frequencies around the first resonance. They also do not account the effect of realistic parameters such as the wavelength dispersion. Instead of the approximated equations, we will  derive the characteristic equations in realistic systems [\figref{OPOandCFS}] in order for the stability analysis. We use the equations derived in the previous section [App.~\ref{sec:TransferFunction}]. Once we get the loop gain of the feedback which composes the characteristic equation, we do not need to consider the transfer function for the stability analysis [App.~\ref{sec:Stability}]. We assume the symmetry around the carrier frequency [Eq.~\eqref{eq:DefSymmetry}] in the case of free space optics, which leads to simplifying the analysis.

\subsection{DOPO -  Positive feedback system}
\label{sec:DOPO}
DOPO can be considered as one of the quantum coherent positive-feedback systems.
The transfer function of DOPO derived in the previous section [Eqs.~\eqref{eq:PSATransferFunction1}-\eqref{eq:PSATransferFunction3}] does not include the phase-matching bandwidth caused by the wavelength dispersion of the refractive index. In order for the precise analysis, we use another transfer function of DOPO [$G_a^{(o)}(\omega ) $ and $G_c^{(o)}(\omega )$], which consists of the transfer function of degenerate optical amplification in a nonlinear material [$G_a^{(s)}(\omega ) $ and $G_c^{(s)}(\omega )$]. According to Ref. \cite{Boyd08}, the transfer function of the process under the pump with the frequency of $2\omega_c$ is given as follows:
\begin{align}
\label{eq:ClassicalParame1}
	A^{(s)}(\omega ) &= G^{(s)}_a (\omega)  A(\omega )  + G^{(s)}_c (\omega) A^*(-\omega ), \\
\label{eq:ClassicalParame2}
	G^{(s)}_a  (\omega) & = \ee^{i\Delta k_\omega l_c/2}\left[ \cosh g_\omega - \frac{i\Delta k_\omega l_c }{2g_\omega} \sinh g_\omega \right], \\
\label{eq:ClassicalParame3}
	G^{(s)}_c (\omega) & = \ee^{i\Delta k_\omega l_c/2} \frac{\kappa_\omega}{g_\omega} \sinh g_\omega,
\end{align}
where
\begin{align}
\label{eq:DeltaKomega}
	\Delta k_\omega &= k_{2\omega_c} - k_{\omega_c + \omega} - k_{\omega_c - \omega}, \\
\label{eq:gomega}
	g_\omega & = \sqrt{\kappa_\omega \kappa^*_{-\omega} -\left(\frac{\Delta k_\omega l_c }{2}\right)^2}, \\
	\kappa_{\pm \omega} &\propto i A_{2\omega_c} l_c \frac{\omega_c \pm \omega}{n_{\omega_c \pm \omega}}.
\end{align}
Here, $k_\omega = n_\omega \omega/c$ and $n_\omega$ are the wave vector and the refractive index at the frequency $\omega$, $c$ is the speed of light in vacuum, $l_c$ is the interaction length, and $A_{2\omega_c}$ is the amplitude of the pump beam. Note that we change the notations of Eq.~(2.9.2) in Ref.~\cite{Boyd08} as $gl\to g_\omega,\ \kappa_1 l \to \kappa_\omega,\ \kappa_2 l \to \kappa_{-\omega},\ \omega_1 \to \omega_c + \omega,\ \omega_2 \to \omega_c - \omega,\ \omega_3 \to 2\omega_c$. The parametric gain $g_\omega$ is determined by the wave vector mismatch $\Delta k_\omega$, $A_{2\omega_c}$, $l_c$, and the second-order nonlinear susceptibility which is omitted since we assume it as a constant for the frequency. The transfer functions $G^{(s)}_a (\omega)$ and $G^{(s)}_c (\omega)$ do not satisfy the the symmetry around the carrier frequency [$\big(G^{(s)}_a (-\omega)\big)^*\neq G^{(s)}_a (\omega)$, $\big(G^{(s)}_c (-\omega)\big)^*\neq G^{(s)}_c (\omega)$], despite  some symmetries: $g_\omega = g_{-\omega}$, $\Delta k_\omega = \Delta k_{-\omega}$.
Under the realistic assumption of perfect phase-matching at the carrier frequency ($\Delta k_{\omega = 0}=0$), 
the wave vector mismatch $\Delta k_\omega$ can be written as 
\begin{align}
\label{eq:DeltaKOmega}
	\Delta k_\omega &= \frac{1}{c}\left[2\omega_c n_{\omega_c} - (\omega_c + \omega)n_{\omega_c +\omega} -   (\omega_c - \omega) n_{\omega_c -\omega}\right].
\end{align}
We also assume the phase of the pump beam as $i A_{2\omega_c} = |A_{2\omega_c}|$ leading to $\kappa_{\pm \omega} \in \mathbb{R}$.
We note that Eq.~\eqref{eq:ClassicalParame1} does not include the phase factor for the propagation of the interaction length with $l_c$. We involve it in another transfer function describing the propagation between controller and plant in the stability analysis.

Now, let us consider a DOPO with the transfer function of parametric amplification [$G_a^{(s)}(\omega ) $ and $G_c^{(s)}(\omega )$]. The loop gains for DOPO and the characteristic function are written as
	\begin{align}
	\label{eq:KaOPO}
		\Lambda_a^{(o)} (\omega ) &= \sqrt{R_{o}(1-L_{o})} \ee^{i \omega n_{\omega_c+\omega}l_{o}/c}G_a^{(s)}(\omega), \\
	\label{eq:KcOPO}
		\Lambda_c^{(o)} (\omega ) &= \sqrt{R_{o}(1-L_{o})} \ee^{i \omega n_{\omega_c+\omega}l_{o}/c}G_c^{(s)}(\omega), \\
	\notag
		\Lambda_D^{(o)} (\omega ) &= [1-\Lambda_a^{(o)} (\omega )]  [1-(\Lambda_a^{(o)} (-\omega ))^*]  \\
	\label{eq:KDOPO}
	&\qquad - \Lambda_c^{(o)} (\omega )(\Lambda_c^{(o)}(-\omega ) )^*,
	\end{align}
where $l_o$, $R_o$, and $L_o$ are the DOPO cavity round trip length, the energy reflectivity, and the intra-cavity energy loss, respectively, while $n_{\omega_c + \omega }$ is the refractive index of the frequency $\omega_c + \omega$ at the propagation path. The transfer functions regarding input-output relation are calculated from Eqs.~\eqref{eq:TransferFunctionFB} as 
	\begin{align}
	\label{eq:GaOPO}
		G_a^{(o)} (\omega)& = \frac{1}{\sqrt{R_{o}}}\left[ -1+\frac{1-(\Lambda_a^{(o)} (-\omega ))^*}{\Lambda_D^{(o)} (\omega )}(1-R_{o})\right], \\
	\label{eq:GcOPO}
		G_c^{(o)} (\omega)& = \frac{1-R_{o}}{\sqrt{R_{o}}}\frac{\Lambda_c^{(o)} (\omega )}{\Lambda_D^{(o)} (\omega )}.
	\end{align}
For the stability, since DOPO has the maximum gain at the carrier frequency and the feedback is positive, it is enough to find the condition at the carrier frequency.

We describe the parametric gain near the carrier frequency under the perfect phase-matching as
\begin{align}
\label{eq:ApproxParameGain}
g_{\omega \approx 0}\approx |\kappa_{\omega = 0}| \equiv C \xi,
\end{align}
where $C$ is a constant, and $\xi \equiv |A_{2\omega_c}|/|A_{\mathrm{th}}|$  ($0\le \xi <1$) is the pump amplitude normalized by the oscillation threshold amplitude of $|A_{\mathrm{th}}|$.
Equation~\eqref{eq:ClassicalParame1} can be approximated as
\begin{align}
\label{eq:ApproxParame}
	A^{(s)}(\omega ) \approx \cosh (C\xi ) A(\omega) +  \sinh (C\xi ) A^*(-\omega ).
\end{align}
By using this approximation, the characteristic equation for the stability condition of DOPO is given as
	\begin{align}
	\notag
		&\Lambda_D^{(o)} (\omega \approx 0) \approx 
		[1-(\Lambda_a^{(o)} (\omega \approx 0 )+\Lambda_c^{(o)} (\omega  \approx 0))]  \\
\notag
 		&\quad \times[(1-(\Lambda_a^{(o)} (\omega \approx 0)-\Lambda_c^{(o)} (\omega \approx 0 ))] \\
\notag
		&= \left[1-\ee^{C\xi}\sqrt{R_{o}(1-L_{o})} \ee^{i\omega n_{\omega_c}l_{o}/c}\right] \\
		&\quad \times\left[1-\ee^{-C\xi}\sqrt{R_{o}(1-L_{o})} \ee^{i\omega n_{\omega_c}l_{o}/c}\right] = 0.
	\end{align}
Here, we also use the approximation as $(\Lambda_j^{(o)} (-\omega \approx 0 ))^* \approx \Lambda_j^{(o)} (\omega \approx 0 )$, $(j=a,c)$ and $n_{\omega_c + \omega} \approx n_{\omega_c}$. The stability condition of DOPO in this form can be written as
	\begin{align}
	\label{eq:DOPOcondition}
		\ee^{C\xi}\sqrt{R_{o}(1-L_{o})}  <1.
	\end{align}
From this oscillation condition, the constant $C$ can be represented by cavity parameters as
\begin{align}
\label{eq:Ccalc}
	C = \ln \frac{1}{\sqrt{R_{o}(1-L_{o})}}.
\end{align}
Therefore, by using the normalized pump amplitude $\xi$ and the cavity parameters $R_{o}$ and $L_{o}$, we can represent the $\kappa_\omega$ as
\begin{align}
\label{eq:kappaXi}
	\kappa _\omega = \frac{\omega_c + \omega}{\omega_c}\frac{n_{\omega_c}}{n_{\omega_c + \omega}} C \xi.
 \end{align}

\subsection{CFS with DOPO in Free Space - Negative feedback system with the symmetry around the carrier frequency}
\label{sec:CFSf}
We consider CFS involving DOPO as shown in \figref{OPOandCFS}. We also consider utilizing BPF in the case of free space optics. The transfer function of second-order Butterworth BPF is represented as
	\begin{align}
		H(\omega ) = \frac{\omega_{\mathrm{HWHM}}^2}{(i\omega - \omega_1)(i\omega - \omega_2)},
	\end{align}
where $\omega_1 = \omega_{\mathrm{HWHM}} \ee^{i 3/4 \pi}$, $\omega_2 = \omega_{\mathrm{HWHM}} \ee^{i5/4\pi}$, and $\omega_{\mathrm{HWHM}}/(2\pi)$ is the cutoff frequency [we assume it is $100$~GHz as in the main text]. Since the cutoff frequency is narrower enough than the phase-matching bandwidth of the nonlinear optics, we approximate the transfer function of a plant as Eq.~\eqref{eq:ApproxParame} for the stability analysis, which has no frequency dependence. Furthermore, in free space, the wavelength dispersion regarding the propagation is negligible, which corresponds to $n_{\omega_c+\omega}=1$ in Eqs.~\eqref{eq:KaOPO}\eqref{eq:KcOPO}. They lead to the the symmetry around the carrier frequency [Eq.~\eqref{eq:DefSymmetry}] for all transfer functions, which means that the representations are simplified in the form of quadratures.
The resulting loop gain of CFS with DOPO in free space is
\begin{align}
\notag
	\Lambda_q^{(\mathrm{CFS},f)}(\omega )& = -\sqrt{R_{f}(1-L_{f})} \ee^{i \omega l_{f}/c} \\
\label{eq:KxCFSf}
	&  \qquad \times H(\omega ) G_q^{(o,f)}(\omega ), \quad (q=x,p), 
\end{align}
where the approximated transfer function at the DOPO $G_q^{(o,f)}(\omega )$ is written by the loop gains in the DOPO $\Lambda_q^{{(o,f)}}(\omega ) $ as
\begin{align}
\label{eq:Gxof}
	G_q^{(o,f)}(\omega ) &= \frac{1}{\sqrt{R_{o}}} \frac{\Lambda_q^{(o,f)}(\omega)-R_{o}}{1-\Lambda_q^{(o,f)}(\omega)}, \\ 
\label{eq:Lambdaxof}
	\Lambda_q^{{(o,f)}}(\omega ) & =\sqrt{R_{o}(1-L_{o})} \ee^{i \omega  l_{o}/c} G_q^{(s)}(0), \\
	 G_x^{(s)}(0) &= G_a^{(s)}(0)+G_c^{(s)}(0) = \ee^{C\xi},\\
	 G_p^{(s)}(0) &= G_a^{(s)}(0)-G_c^{(s)}(0) = \ee^{-C\xi}.
\end{align}
Here, $C$ is the constant regarding the single pass squeezing level given by Eq.~\eqref{eq:Ccalc}.
The characteristic function in Eq.~\eqref{eq:Dsymmetry} is
\begin{align}
\notag
	\Lambda_D^{(\mathrm{CFS},f)} (\omega ) &= [1-\Lambda_x^{{(\mathrm{CFS},f)}} (\omega )] \\ 
	&\quad \times [1-\Lambda^{{(\mathrm{CFS},f)}} _p (\omega )].
\end{align}
Since $|\Lambda^{{(\mathrm{CFS},f)}} _p (\omega )|<1$, the characteristic equation $\Lambda_D^{(\mathrm{CFS},f)} (\omega =i s) =0$ becomes
\begin{align}
	1-\Lambda_x^{{(\mathrm{CFS},f)}} (\omega =i s )=0.
\end{align}
\subsection{CFS with DOPO in a Waveguide - Negative feedback system without the symmetry around the carrier frequency}
\label{sec:CFSw}
We consider CFS in a waveguide. Unlike CFS in free space, we need to consider the effects of the wavelength dispersion in the waveguide. Thus, we can not simplify the representations in the form of quadratures.
The loop gains of CFS with DOPO in a waveguide are expressed as
\begin{align}
\notag
	\Lambda_j^{(\mathrm{CFS},w)}(\omega )& = -\sqrt{R_{f}(1-L_{f})} \ee^{i \omega n_{\omega_c+\omega}l_{f}/c}\\
	&  \qquad \times G_j^{(o)}(\omega ), \quad (j=a,c).
\end{align}
Note again that the transfer function of DOPO $G_j^{(o)}(\omega )$ is given by Eqs.~\eqref{eq:GaOPO}\eqref{eq:GcOPO}, which also consists of the loop gains of DOPO $\Lambda_j^{(o)}(\omega )$ in Eqs.~\eqref{eq:KaOPO}-\eqref{eq:KDOPO} and the transfer function of the parametric amplification $G_j^{(s)}(\omega )$ in Eqs.~\eqref{eq:ClassicalParame2}\eqref{eq:ClassicalParame3}. The characteristic function [Eq.~\eqref{eq:KD}] is 
	\begin{align}
	\notag
		\Lambda_D^{(\mathrm{CFS},w)} (\omega ) &= [1-\Lambda_a^{(\mathrm{CFS},w)} (\omega )] \\
	\notag
		&\quad \times [1-(\Lambda^{(\mathrm{CFS},w)} _a (-\omega ))^*)  \\
	&\quad - \Lambda_c^{(\mathrm{CFS},w)} (\omega )(\Lambda^{(\mathrm{CFS},w)} _c(-\omega ) )^*.
	\end{align}
The characteristic equation is 
\begin{align}
\label{eq:WG_CE}
\Lambda_D^{(\mathrm{CFS},w)} (\omega ) =0.
\end{align}

\section{Gain Spectrum Obtained from Singular Value}
\label{sec:AppGainSpec}
As we explained in Eq.~\eqref{eq:Gmatrix}, the transfer function matrix in the quadrature form has nonzero off-diagonal elements in general. The transfer function can be decomposed into a rotation and a lossy phase-sensitive amplifier and a rotation for quadratures by using singular value decomposition (called the Bloch--Messiah reduction in the ideal limit without loss \cite{Braunstein05,Furusawa11}) which is expressed as follows:
\begin{align}
\notag
	&
	\bm{J}
	\begin{pmatrix}
		G^{(s)}_a(\omega)  & G^{(s)}_c (\omega ) \\
		(G^{(s)}_c(-\omega ))^*   & (G^{(s)}_a(-\omega))^*
	\end{pmatrix}		
	\bm{J}^{-1} \\
	&\qquad 
	=U(\theta ) 
	\begin{pmatrix}
		G^{(s)}_{\tilde x}(\omega)  & 0 \\
		0  & G^{(s)}_{\tilde p}(\omega)
	\end{pmatrix}
	U(\phi ),\\
	&U(\theta ) \equiv 
	\begin{pmatrix}
		\cos\theta  & -\sin\theta \\
		\sin \theta & \cos\theta
	\end{pmatrix},
\end{align}
where $U(\theta)$ is the matrix representing $\theta$ rotation for $\hat X$ and $\hat P$ quadratures, while $G^{(s)}_{\tilde x}(\omega) $ and $G^{(s)}_{\tilde p}(\omega)$ are the singular values ($G^{(s)}_{\tilde x}(\omega) \ge G^{(s)}_{\tilde p}(\omega) \ge 0$). The larger singular value corresponds to the single pass gain. In the same manner, we denote the larger singular value of the transfer function matrix for DOPO as $G^{(o)}_{\tilde x}(\omega)$.

\end{document}